\documentclass[12pt]{arxiv-article}
\pdfoutput=1

\usepackage{tikz-cd}
\usepackage{definitions}

\usepackage[percent]{overpic}
\usepackage{braket}
\usepackage{caption}

\usepackage{simeontex}

\usepackage{physics}

\usepackage{cite}

\usepackage{mdframed}
\usepackage{tablefootnote} 
\makeatletter 
\AfterEndEnvironment{mdframed}{%
 \tfn@tablefootnoteprintout%
 \gdef\tfn@fnt{0}%
}
\makeatother

 \usepackage[textsize=small]{todonotes}
 \RequirePackage{framed}

\newcommand{\qed}{\hfill\ensuremath{\square}}
\newtheorem{statement}{Statement}

\usepackage{adforn}

\newcommand*{\nmax}{{\Delta n_{\text{max}}}}
\newcommand*{\cH}{\mathcal{H}}
\newcommand*{\cC}{\mathcal{C}}
\newcommand*{\cQ}{\mathcal{Q}}
\newcommand*{\LCFT}{\Lambda_{\textsc{cft}}}
\newcommand*{\Hilb}{\mathfrak{H}}
\newcommand*{\Op}{\mathcal{O}}

\preprint{\texttt{IPMU17-0148}}

\newcommand{\OfficialTitle}{Quantum Information Theory \\ of 
the Gravitational Anomaly
}

\title{%
  \Huge
  \textbf{
      \dosserif
      \OfficialTitle
    }  
}

\hypersetup{pdfauthor={Simeon Hellerman and Domenico Orlando and
     Masataka Watanabe},pdftitle={\OfficialTitle}}

\author{%
  \begin{minipage}{1\linewidth}
    \begin{center} \dosserif
      {\small 
        \textbf{Simeon Hellerman}\textsuperscript{1}, 
        \textbf{Domenico~Orlando}\textsuperscript{1,2,3} and
        \textbf{Masataka~Watanabe}\textsuperscript{1,3,4,5}}
    \end{center}
    \vspace{0.2cm}
\setstretch{1.1}
    \authorBlock{1}{Kavli Institute for the Physics and Mathematics of the Universe (\textsc{wpi}).\\ The University of Tokyo.\\Kashiwa, Chiba 277-8582, Japan}
    \authorBlock{2}{\textsc{infn} sezione di Torino. Via Pietro Giuria 1, 10125 Torino, Italy}
    \authorBlock{3}{Albert Einstein Center for Fundamental Physics, \textsc{itp}, University of Bern. \\Sidlerstrasse 5, \textsc{ch}-3012 Bern, Switzerland}
    \authorBlock{4}{Department of Physics, Faculty of Science,
University of Tokyo, \\Bunkyo-ku, Tokyo 133-0022, Japan}
\authorBlock{5}{Department of Particle Physics and Astrophysics,\\ Weizmann Institute of Science, Rehovot 7610001, Israel}
  \end{minipage}
}

\newcommand{\mw}[1]{\textcolor{olive}{Note: #1 -- MW}}
\newcommand{\imp}[1]{\textcolor{black}{#1}}

 \definecolor{anti-flashwhite}{rgb}{0.95, 0.95, 0.96}
  \definecolor{White}{rgb}{1,1,1}

{\color{Purple}}%
{\color{Black}}
{\color{Purple}}%
{\color{Black}}
{\color{Red}}%
{\color{Black}}
\def\checkmark{\tikz\fill[scale=0.4](0,.35) -- (.25,0) -- (1,.7) -- (.25,.15) -- cycle;} 
\def\ok{\green{\checkmark}}
\def\ppq{\purple{{    \scalebox{1.5}{??} }}}
\def\blut{\blue{{    \scalebox{1.5}{?} }}}

\makeatletter
\DeclareRobustCommand\tot{\mathbin{\mathpalette\frown@otimes\relax}}
\newcommand{\frown@otimes}[2]{%
  \vbox{
    \ialign{##\cr
      \hidewidth$\m@th#1{}_\frown$\kern-\scriptspace\hidewidth\cr
      \noalign{\nointerlineskip\kern-1pt}
      $\m@th#1\otimes$\cr
    }%
  }%
}
\makeatother

\def\efac{\e\lrm f}

\date{}

\def\Id{{\bf 1}}

\def\MacroForIOrA{I}
\begin{document}

\begin{titlepage}
\setstretch{1.1}

  \maketitle
  \thispagestyle{empty}

  \vfill
  \setstretch{1}
  \abstract{
  \normalfont \noindent
    We show that the standard notion of entanglement is not defined for gravitationally anomalous two-dimensional theories  because they do not admit a local tensor factorization of the Hilbert space into local Hilbert spaces.  Qualitatively, the modular
    flow cannot act consistently and unitarily in a finite region, if there are different numbers of states with a given energy traveling in the two opposite directions.  We
    make this precise by decomposing it into two observations:
    First, a two-dimensional \acl{cft} admits a consistent quantization on a space with boundary only if it is not anomalous.
    Second, a local tensor factorization always leads to a definition of consistent, unitary, energy-preserving boundary condition.
    As a corollary we establish a generalization of the Nielsen--Ninomiya theorem to all two-dimensional unitary local \aclp{qft}:
    No continuum quantum field theory in two dimensions can admit a lattice regulator unless its gravitational anomaly vanishes.
We also show that the conclusion can be generalized to six dimensions by dimensional reduction on a four-manifold
of nonvanishing signature.  We advocate that these points be used to reinterpret the  
gravitational anomaly quantum-information-theoretically, as a fundamental obstruction to the localization of quantum information.
}

\end{titlepage}

\restoregeometry

\setstretch{1.15}

\setcounter{tocdepth}{2}
\setcounter{secnumdepth}{2}
\tableofcontents

\newpage

\section{Introduction and statement of results}
\label{sec:introduction}

\heading{Quantum information, tensor factorization and the gravitational anomaly}

Much progress has been made recently in the understanding of quantum gravity using ideas from quantum information theory
via the holographic duality with \ac{cft}.
  These advances~\cite{Ryu:2006bv,Maldacena:2013xja,Bousso:2014sda,Almheiri:2014lwa,Maldacena:2015waa,Balakrishnan:2017bjg} have relied on commonly-used notions in quantum information theory, defining entanglement between subsectors
  of the \ac{cft} in terms of a tensor decomposition of the system into factors describing Hilbert spaces supported in
  complementary regions of a spatial slice.  Many ideas and intuitions surrounding such factorizations come
  from the idea of regulating the system by placing it on a spatial lattice or other discretization of space.

However, the existence of a lattice regulator is \rwa{not} a
universal or even a generic property of \aclp{qft}.  Indeed, many familiar continuum theories, particularly those involving chiral fermions,
do not admit a lattice regulator of any kind~\cite{Nielsen:1980rz,Nielsen:1981xu,Nielsen:1981hk}.
Because of this, all lattice realization of chiral theories have required to modify the theory and for example break explicitly chiral symmetry as in~\cite{Ginsparg:1981bj,Shamir:1993zy}, or double the number of fermions~\cite{Kogut:1974ag}.
The obstruction to the existence of a local lattice regulator should be thought of as an \rwa{intrinsic property} of certain
\rwa{continuum} theories.  This property is by definition preserved by \ac{rg} flow %
and is intimately
connected with various anomalies.

One is therefore led to
  wonder whether intuitive notions of quantum information motivated from lattice field theories, may also
  apply less than generally.  In this note we answer this question in the affirmative: \imp{The notion of tensor decomposition 
  of a Hilbert space into factors supported in complementary regions of space, does not apply to theories in
  which certain pure gravitational anomalies are nonzero}.
  
\heading{Subtlety 1: Precise meaning of tensor factorization in continuum QFT}  

When we talk about tensor decomposition of Hilbert spaces,
one could either be talking about Hilbert space factorization in the literal mathematical sense, or in the approximate sense in terms of the Hilbert space of a regulated \ac{qft}.
Quite importantly, it is in the latter sense that Hilbert space factorization is meant in the present paper.
As we shall explain, this definition will be independent of how one defines the theory in the UV, and is only governed by the universality.

Indeed, there are already various known \emph{short-distance} obstructions to tensor factorization in the former, literal sense~\cite{Casini:2013rba,Radicevic:2014kqa,Huang:2014pfa,Radicevic:2015sza,Harlow:2015lma}.
These obstructions are associated strictly with short-distance physics near the entangling surface; the presence and nature of such obstructions depend entirely on the details of the UV-completion, rather the long-distance physics in the continuum limit.
Such literal obstructions to tensor factorization can be resolved by slightly generalizing the notion of tensor product by (1) adding localized degrees of freedom at the entangling surface,\footnote{In the usual terminology "entangling surface" is the boundary of a subregion of a spatial slice, so codimension-two in the full $D$-dimensional
geometry.  In most of the paper we will be considering the case $D=2$, in which the "entangling surface" is zero-dimensional.} and/or by (2) subtracting local degrees of freedom with additional gauge symmetries or constraints that act only near the entangling surface.

In the present paper, we shall absorb all such short-distance ambiguities into a generalized notion of a tensor product, $\tot$, which allows for such local modifications
of the relationship between the Hilbert spaces on ${\cal H}\ll A, {\cal H}\ll B, $ and ${\cal H}\ll {A\cup B}$ that leave the operator algebra undisturbed in the interior of
regions $A,B$ but may modify the algebra near the entangling surface.
We will refer to this kind of modified relationship as a ``renormalized tensor product''; as such renormalized quantities should be, this notion is meant to be insensitive to the UV-completion of the continuum \ac{qft}.\footnote{In fact, we should think of this concept as being precisely define by $\mathcal{F}[\mathcal{M}]$ defined below.}
An illustrative example of this renormalized tensor product is given in~\cite{Harlow:2015lma}, where one introduces heavy fields to recover literal factorization of the gauge theory Hilbert space.
Note that, as explained in~\cite{Harlow:2015lma}, if such a procedure is applied to a bulk gravity theory, one gets nontrivial and striking consistency conditions on the dynamics of Einstein-Maxwell theory \emph{via} gauge-gravity duality.

The obstruction to tensor factorization in this paper refers not to aforementioned short-distance obstructions to literal factorization.
Rather, the anomaly-related obstruction to the renormalized tensor structure is truly a long-distance phenomenon.
Let us stress once more that this is entirely distinct and qualitatively different from the short-distance counterpart;
This obstruction cannot be cured by any short-distance modifications of the continuum theory.
This can be seen because, as we will prove in the main body of the paper, anomalies serve as an obstruction to renormalized tensor factorization.
Since the anomaly is an RG-invariant object, there is no way in which one can cure this type of obstruction by modifying observables or dynamics at short distances.

\heading{Subtlety 2: Precise meaning of partial traces and the locality criterion}

Now, let us state what we precisely mean by partial traces in continuum \ac{qft}.
Hereafter we restrict our attention to two-dimensional theories, as we will only use them in the main body of the present paper.
This simplifies the situation a lot as entangling surfaces are just entangling points in this case.

Seldom talked about explicitly, but advocated correctly in~\cite{Ohmori:2014eia} (which we will also discuss in Section \ref{sec:related}), there are no canonical ways in which to relate $\mathcal{H}_A\ot \mathcal{H}_B$ into $\mathcal{H}_{A\cup B}$.
One needs to choose one such map $\mathcal{M}$ in order to define the partial trace
\begin{equation}
    {\cal M}: \mathcal{H}_{A\cup B}\to \mathcal{H}_A\ot \mathcal{H}_B,
\end{equation}
which is required to define \emph{e.g.} the entanglement entropy.
We call this a factorization map (not \emph{the} factorization map because there are arbitrary choices of such maps~\cite{Ohmori:2014eia}).
After all these procedures, we can finally define the partial trace of a density matrix $\rho$ as 
\begin{equation}
    \rho_A\equiv\Tr_B\left(\mathcal{M}\rho \mathcal{M}^\dagger\right)=
    \Tr_B\left(\mathcal{M}^\dagger\mathcal{M}\rho\right)
\end{equation}
where now $\Tr_B$ can be understood literally as a sum over matrix elements of $\mathcal{M}^\dagger\mathcal{M}\rho$ between (normalizable) states $\Ket{\Psi_i^{(B)}}\in \mathcal{H}_B$.
Notice that 
\begin{equation}
    \mathcal{F}[\mathcal{M}]\equiv\mathcal{M}^\dagger\mathcal{M}:\mathcal{H}_{A\cup B}\to\mathcal{H}_{A\cup B}
\end{equation}
needs to have finite matrix elements to make sense of the above expression, and hence it cannot be an exactly local operator (although it is true that such an operator is centered around the entangling surface).

This can intuitively read off as a modification of the density matrix around the entangling surface using $\mathcal{F}[\mathcal{M}]$,
\begin{equation}
\rho_A=
    \Tr_B\left(\rho_\mathcal{M}\right),\quad \rho_\mathcal{M}\equiv\mathcal{F}[\mathcal{M}]\cdot\rho.
\end{equation}
In order to capture the correct physical correlation between two subregions $A$ and $B$, the effect of $\mathcal{F}[\mathcal{M}]$ should not introduce any correlations of $O(1)$ that were absent previously in the original theory, and
we therefore demand the factorization map be local.
By this we mean $\mathcal{F}[\mathcal{M}]$ needs to be as local as possible, but again we cannot take it exactly local.
Hence, one is forced to introduce a tiny but non-zero factorization timescale $\epsilon_{\rm f}$, which characterizes the time resolution of the factorization map.
The locality criterion for the factorization therefore is that the map $\mathcal{F}[\mathcal{M}]$ commutes (up to subleading corrections in $O(\epsilon_{\rm f})$) with any operators outside the disk of size $\epsilon_{\rm f}$ centered on the entangling surface.

The properties of the factorization map ${\cal M}$ give a precise meaning to the renormalized tensor product $\tot$ without assuming the existence of
a boundary condition or referring to any cutoff on the \ac{cft} itself.
The operation $\mathcal{F}[\mathcal{M}]$ itself is some almost-local operator of size $\epsilon_f$, in which we have some nonuniversality.  This nonuniversality indicates that the renormalized tensor product is not unique for a given QFT and decomposition of the space into regions, but depends on the nature of the physical operation at entangling surface defining the factorization map, as we will see in the rest of the paper.  In
the case where the operation ${\cal M}$ is defined by a boundary condition, as in \cite{Nishioka:2015uka} or the earlier part of \cite{Ohmori:2014eia}, the tensor product depends on the boundary condition; in
the case ${\cal M}$ is inherited from a bulk UV cutoff, as in the later part of \cite{Ohmori:2014eia}, the renormalized tensor product depends on the 
details of the cutoff; and in general it may correspond to some other unspecified procedure altogether, so long as that procedure is defined by a product of almost-local operators
defined at the entangling surface.  We make no further assumption on the nature of ${\cal M}$ at all beyond the almost-locality property of ${\cal F}[{\cal M}]$.  The existence
of a ${\cal M}$ with this almost-locality property implies that ${\cal H}\ll {A\cup B} = {\cal H}\ll A \tot {\cal H}\ll B$ in the sense of the renormalized tensor product discussed earlier.

We will show that there are cases in which there are no such almost-local operators in the theory, meaning that there is no notion of local tensor factorization in that specific theory.  
Our proof of the nonexistence of such a ${\cal M}$ in the presence of a nonvanishing anomaly will be general enough to rule out any local prescription whatsoever for defining the factorization.

\heading{Statement of results and logical flow of the proof}

Now we are ready to state clearly the statement of the main result and the assumptions behind it: \imp{
A relativistic two-dimensional \ac{qft} with a non-vanishing gravitational anomaly, \emph{i.e.}, $c_L\neq c_R$, defined on $\Sigma=A\cup B$ does not admit a Hilbert space factorization $\mathcal{H}_{A\cup B}=\mathcal{H}_A\tot \mathcal{H}_B$, which is local \textcolor{black}{({\it in the sense we explained in Subtlety 2 above})}, even in any generalized sense of tensor factorization \textcolor{black}{({\it allowing for local short-distance modifications at the entangling surface, as explained in Subtlety 1 above.})}.
}

We are going to prove the statement in minimal logical steps as follows.
\begin{description}
\item[Step 1] The existence of tensor factorization of Hilbert space (again in the sense that we explained in detail above) implies a consistent, unitary quantization of the conformal field theory on a space with boundary (with finite boundary entropy).
\item[Step 2] Conformal field theories on the boundary (with finite boundary entropy) always have a vanishing gravitational anomaly.
\item[Step 3] Combining Step~1 and~2, the existence of tensor factorization implies a vanishing gravitational anomaly.
\end{description}
Finally taking the contrapositive, we prove the promised statement.

Schematically, the flow of the logic is as follows,
\begin{center}
\begin{tikzcd}
  \text{\fbox{existence of Hilbert space factorization}}\arrow[Rightarrow]{d}\\
  \text{\fbox{consistent boundary condition}}\arrow[Rightarrow]{d}\\
  \text{\fbox{vanishing gravitational anomaly}}
  \end{tikzcd}
\end{center}
and taking the contrapositive, we prove the statement above.
\begin{center}
\begin{tikzcd}
  \text{\fbox{gravitational~anomaly}}\arrow[Rightarrow]{d}\\
  \text{\fbox{obstruction~to~quantization~with~boundary}}\arrow[Rightarrow]{d}\\
  \text{\fbox{obstruction~to~Hilbert~space~factorization}}
  \end{tikzcd}
\end{center}
Also, by using Zamolodchikov's $c$-function, we can trivially see that the result applies not only for \ac{cft}s, but for \ac{qft}s also.

Slightly modifying the above proof leads to a different but equally important result.
By noticing that having a lattice regulator for a continuum theory naturally leads to a consistent, unitary quantization of such a theory on a space with boundary, we shall prove the following statement:
\imp{
Any gravitationally anomalous relativistic two-dimensional \ac{qft}s (including strongly-coupled as well as non-Lagrangian theories) do not admit the existence of lattice realizations with finite entropy per site.
}
We also show below the schematic logical flow of the proof (or rather, the contrapositive of it):
\begin{center}
\begin{tikzcd}
& \textcolor{Gray}{\text{\fbox{existence of Hilbert space factorization}}}\arrow[Rightarrow,gray]{d}\\
  \text{\fbox{existence of lattice regulator}}\arrow[Rightarrow]{r} &
  \text{\fbox{consistent boundary condition}}\arrow[Rightarrow]{d}\\
   & \text{\fbox{vanishing gravitational anomaly}}
  \end{tikzcd}
\end{center}

This gives a broad generalization of the (two-dimensional) Nielsen-Ninomiya theorem~\cite{Nielsen:1980rz,Nielsen:1981xu,Nielsen:1981hk} that is applicable to general unitary interacting \ac{qft}s in two dimensions, while the very original proof was only applicable to free fermionic systems.
Note that this observation is also in line with the recent
interest in symmetry-protected topological phases/states, studied both by high-energy and condensed matter theorists, in that it also tells us the relation between anomaly matching and the existence of a symmetry-obeying regulator~\cite{Kravec:2013pua,Wang:2013yta}.

Finally, generalisations to six-dimensional theories are possible, and using dimensional reduction we prove that \imp{
a six-dimensional \ac{qft} with non vanishing factorized term in the gravitational anomaly polynomial does not have Hilbert space tensor factorization along a (four-dimensional) entangling surface with non-zero signature class. 
}

\heading{Holographic implications}

This has also interesting holographic implications. 
For example, because of the main result of the present paper, it turns out that the recent general proof of the \ac{qnec}~\cite{Balakrishnan:2017bjg} makes assumptions about localizability of quantum information that are apparently less than general.
This is not an abstract problem: there do exist known examples of \ac{uv}-complete \(AdS_3 / CFT_2\) holographic dual pairs with large gravitational anomalies with
\bbb
1 \muchlessthan \abs{ C_L - C_R} \muchlessthan C_L + C_R\ ,
\een{HolographicAnomalyHierarchy}
We shall mention examples briefly in Section \ref{sec:holography-and-anomaly}.

These gravitational theories, which are in the holographic regime 
\bbb
L\lrm{AdS} \muchgreaterthan m\uu{-1}\ll{{{\rm heavy}\atop{\rm states}}} \cc
\gtrsim L\lrm{planck} 
\eee
 admit no local definition of entanglement at all, even
modulo the usual UV-subtleties.
The existence of these theories should force
a more careful re-examination the question
of what quantities exactly the prescription of~\cite{Ryu:2006bv}
is actually calculating in those cases.  We do not resolve this puzzle in the
present paper, though.

\heading{Outline of the paper}

The outline of the paper is as follows.
In Section~\ref{sec:lattice-implies-boundary} we show how the existence of a lattice regulator for a two-dimensional \ac{cft} implies the existence of a conformal \acl{bc}. 
In Section~\ref{sec:boundary-implies-no-anomaly} we see that a non-trivial \ac{bc} is only possible for theories with no gravitational anomaly and, more in general, when continuous symmetries are present, only for theories where the holomorphic and anti-holomorphic anomaly coefficients coincide.
This gives an immediate generalization of the Nielsen--Ninomiya theorem to any two-dimensional \ac{cft} and, using Zamolodchikov's \(c\)-function, to any two-dimensional \acl{qft} (Section~\ref{sec:generalized-NN-theorem}).
In Section~\ref{sec:factorization-implies-boundary} we refine the first implication and show that the tensor factorization of the Hilbert space implies a consistent \ac{bc} for a two-dimensional \ac{qft}.
This implies immediately that there is no usual notion of entanglement for anomalous two-dimensional \acp{qft} (Section~\ref{sec:quantum-information}).
We generalize this result via dimensional reduction in Section~\ref{sec:six-dimensions} to show that a  six-dimensional theory whose factorized anomaly coefficient is not vanishing cannot have entanglement across a surface with non-zero signature class.
In section~\ref{sec:discussion} we discuss some implications of our results from the point of view of \acl{qft} and holography.

For the convenience of the reader, we partly summarized the above into a logical flow diagram and indicated how each step is proven in each section.
\begin{center}
\begin{tikzcd}
& {\text{\fbox{existence of Hilbert space factorization}}}\arrow[Rightarrow, "\text{Sec. \ref{sec:factorization-implies-boundary}}", shift left=1.5ex]{d}\\
  \text{\fbox{existence of lattice regulator}}\arrow[Rightarrow,"\text{Sec. \ref{sec:lattice-implies-boundary}}"]{r} \arrow[Rightarrow, "\text{trivial}"]{ur}&
  \text{\fbox{consistent boundary condition}}\arrow[Rightarrow, "\text{Sec. \ref{sec:boundary-implies-no-anomaly}}"]{d}\arrow[Rightarrow,shift left=1.5ex,"\text{Sec. \ref{ApproxFacStateConstruction}}"]{u}\\
   & \text{\fbox{vanishing gravitational anomaly}}
  \end{tikzcd}
\end{center}

\section{Gravitational anomaly obstructs lattice regularization in two-dimensional QFTs}

\subsection{QFTs with lattice regulators admit boundary conditions}
\label{sec:lattice-implies-boundary}

The first logical step is to prove the following assertion.

\begin{statement}
  \label{lemma:lattice-implies-boundary}
If a two-dimensional \ac{cft} \(\cC\) admits a local lattice regulator then there exists a unitary, energy-preserving conformal \acl{bc} for \(\cC\).
\end{statement}

\heading{Lattice Hamiltonian with bounded interactions}

Let \(\cC\) be a two-dimensional \ac{cft} with a \ac{uv} completion described by a Hamiltonian system living on a lattice.
This means that the fundamental \aclp{dof} \(\phi_m\) live on the sites of a lattice with spacing $\e\ll{\textsc{uv}} =\L\ll{\textsc{uv}}\uu{-1}$, labeled by an integer \(n \in \setZ\).
We assume that the system has finite entropy per site.
We also assume that the theory is local in the sense that the \acp{dof} satisfy local commutation relations and that the Hamiltonian \(\cH\) only couples \ac{dof} separated by a bounded number \(\nmax \) of lattice sites.
Then the Hamiltonian is a sum of local contributions
\begin{equation}
  \cH =\e\ll{\textsc{uv}}\cc \sum_{n = - \infty}^{\infty} \cH_n \ ,
\end{equation}
where $\e\ll{\textsc{uv}} = \L\ll{\textsc{uv}}$ is the lattice spacing.
Each local contribution to the Hamiltonian \(\cH_n\) is an operator coupling to  \ac{dof} at sites $n\pr$ a bounded
number of lattice sites $n - n\pr$ away:\footnote{
The form of the Hamiltonian is called ultra-local, but the proof below is the same for local Hamiltonians, too.
}
\begin{equation}
  \cH_n = \mathcal{F}_n( \phi_{n'}\cc {\tt s.t.}\cc\cc \abs{n - n'} \le \nmax ) .
  \label{EquationAboveWhereWeMentionTheCFTScaleLambdaCFT}
\end{equation}

Of course a generic lattice Hamiltonian does not flow in the continuum to a \ac{cft} unless the couplings are finely tuned.
We assume that this is the case for \(\mathcal{H}\), that by hypothesis flows at a scale \(\LCFT\) to the theory \(\cC\) of interest.

\heading{Construction of the boundary Hamiltonian}

How does the existence of \(\cH\) imply a consistent unitary quantization of \(\cC\) on a space with boundaries? The idea is to start with \(\cH\) and delete all the \ac{dof} living on the negative lattice sites.
To be more precise, we define a new Hamiltonian \(\cH^+\) as
\begin{equation}
  \cH^+ = \e\ll{\textsc{uv}}\cc \sum_{n = 0}^{\infty} \cH_{n + \nmax} .\llsk
\end{equation}
It is immediate to see that \(\cH_{n + \nmax}\) has support only on lattice points on the right of \(n\) and that the new Hamiltonian \(\cH^+\) only contains interactions among lattice points in \(n \ge 0\). No special values of the \ac{dof} are required in \(n = 0\), \emph{i.e.} \(\cH^+\) has free boundary conditions in \(n = 0\) (this is well-defined because we have a specified lattice regulator).
Where does \(\cH^+\) flow to at \(\LCFT\)?
Remember that the local Hamiltonians \(\cH_n\) have support bounded by \(2\nmax\).
It follows that sufficiently away from the boundary at \(n = 0\), the local dynamics of the theory is the same for \(\cH^+\) and \(\cH\).
The region of size \(\nmax\) that is affected by the deletion of the negative lattice sites goes to zero in the continuum limit and gives rise to a boundary condition.
All in all, \(\cH^+\) flows to the same theory as \(\cC\), but this time defined on \(\setR^+\), with some \ac{bc} that are to be determined at the origin.
Unitarity and energy conservation are manifest on the lattice theory on \(\setN_0\) and are inherited by the continuum theory on \(\setR^+\) at the scale \(\LCFT\).  

In order to complete the proof, we also need to see that the boundary condition in the IR becomes conformal, but
it is less obvious that this is so --
The way we define the boundary condition in the UV is obviously not conformal.
However, for a sufficiently
large hierarchy between the infrared scale and the lattice
scale, the boundary condition can be shown under a mild
hypothesis to flow to a conformal boundary condition (we refer to this as the ``Friedan--Konechny hypothesis''~\cite{Friedan:2003yc,Friedan:2005dj}).
Under this assumption, we have proven that the existence of a lattice regulator \(\cH\) for a two-dimensional \ac{cft} \(\cC\) defined on \(\setR\), implies the existence of a \ac{bcft} \(\cC^+\) defined on \(\setR^+\).
All the more, such a boundary conformal field theory has finite boundary entropy, because of the assumption of finite entropy per site, admitting the existence of the boundary state characterizing the boundary condition.\qed

Since the Friedan--Konechny hypothesis is crucial to our argument, we would like to discuss it more in detail.
At the scale $\L\lrm{CFT}$ there will in general still be nonconformal operators.
The intuition is that if such operators are irrelevant then they will flow to zero at sufficiently long distances and the \ac{bc} will flow to a conformal \ac{bc}; if such operators are relevant, on the other hand, they will give large frequencies to some \ac{dof} that can be integrated out to give a \ac{bc} with fewer \ac{dof}. Eventually this process will terminate reaching a (possibly trivial) fixed point and hence a conformal \acl{bc}, at a scale one can call $\L\lrm{{{\rm conformal}\atop{\rm boundary}}}$.
This intuition is made more precise in Friedan and Konechny~\cite{Friedan:2003yc} in terms of a functional on boundary theories on two-dimensional \acp{cft}, the Affleck--Ludwig \(g\)-function~\cite{Affleck:1991tk}, that, in analogy with Zamolodchikov's \(c\)-function is monotonic along the \ac{rg} flow and stationary if and only if the flow is at a fixed point.
At the time of writing it is not known if \(g\) is bounded below, but we assume that this is the case. %

\subsection{Boundary consistency implies vanishing anomaly}
\label{sec:boundary-implies-no-anomaly}

\begin{mdframed}[%
  skipabove=.5cm,
  leftmargin=.5cm,%
  rightmargin=.5cm,%
  innertopmargin=.25cm%
  innerbottommargin=1.25cm
  ]
{\it The results in this section originally appeared in the Master's thesis of one of the authors~\cite{WatanabeThesis:2015}.}\tablefootnote{During the preparation of this paper we learned of related work by K.~Jensen, E.~Shaverin, and A.~Yarom~\cite{Jensen:2017eof}, which beautifully transcribes the no-go theorem for
boundary conditions discovered originally in~\cite{WatanabeThesis:2015}, into the language of anomaly polynomials, confirming the conclusion of~\cite{WatanabeThesis:2015} and drawing
a conclusion opposite to that of the earlier preprint~\cite{Nishioka:2015uka}. In connection with the latter, the
authors are grateful to T. Nishioka for many stimulating
questions when we first presented these results publicly in their current form~\cite{SHSeminarsIncludingIPMU2015}, and for correspondence afterwards between ourselves
and the authors of~\cite{Nishioka:2015uka} on the equivalence of their results to those of~\cite{WatanabeThesis:2015} which appeared earlier.}
\end{mdframed}

\heading{Existence of conformal boundary states implies $c_L=c_R$}

We have seen that theory with a lattice regulator admits a conformal boundary state.
Now we would like to show that this implies in turn that the theory cannot have a gravitational anomaly.
\begin{statement}
  \label{lemma:boundary-implies-no-anomaly}
  A  two-dimensional \ac{cft} admits a non-trivial boundary state \(\ket{B}\)  only if \(C_L = C_R \).
\end{statement}

Let \(\ket{B}\) be a boundary state representing an energy-conserving conformal \ac{bc}.
Then the boundary condition \(T(z) = \tilde T(\bar z)\) for the energy-stress tensor implies that~\cite{Cardy:1984bb,Cardy:1986gw,Cardy:1989ir,Cardy:2004hm}%
\footnote{Our proof applies to more general boundary conditions.
  Conservation of energy allows for \(T(z) - \tilde T(\bar z)=\partial_0 \Op_{\text{boundary}}\), but also in this case \eqref{eq:Cardy-boundary-condition} holds  modulo terms of the type $\comm{\Op_{\text{boundary}}}{\Op_{\text{boundary}}}$ that cannot close
  on another conformally invariant term.  This is because operator  $\Op_{\text{boundary}}$ must be truly marginal, as it must be if it appears in the boundary Hamiltonian.   See for
  instance section~3.1 of~\cite{Recknagel:1998ih}. }
\begin{align}
\label{VirasoroBoundaryCondition}
  \pqty{L_n - \tilde L_{-n}} \ket{B} = 0  && \forall n .
\end{align}
Acting on this condition with \(\pqty{- L_{-n} + \tilde L_n}\) and using the Virasoro algebra commutation relations we obtain:
\begin{align}
  \label{eq:Cardy-boundary-condition}
  \bqty{ 2n \pqty{L_0 - \tilde L_0} + \frac{n \pqty{n^2 -1}}{12} \pqty{ C_L - C_R }} \ket{B} = 0 && \forall n .
\end{align}
Combining \rr{eq:Cardy-boundary-condition} 
with the condition \rr{VirasoroBoundaryCondition} on \(\ket{B}\) at \(n = 0 \) and
we find that
\begin{equation}
  \pqty{C_L - C_R } \ket{B } = 0.
\end{equation}
This proves statement~\ref{lemma:boundary-implies-no-anomaly}, that a non-trivial boundary state is only possible if the theory has no gravitational anomaly.\qed

\heading{Symmetry preserving boundary and vanishing of the anomaly}

The same condition on the vanishing of the anomaly has to be satisfied also when there is a boundary state that preserves a continuous symmetry group \(G\).
To see that, let \(J^a(z)\) be the corresponding holomorphic Noether current.
The current-current \ac{ope} is given by
\begin{equation}
  J^a(z)J^b(0)
  = - \frac{k_L }{4\pi^2 z^2} g^{ab} + \frac{{f^{ab}}_c}{2\pi z} J^c(0) + \text{(non-singular terms)},
\end{equation}
where $k_L$ is the anomaly coefficient, $g^{ab}$ is the group metric of $G$,
and $f^{abc}$ are the structure constants of the algebra of $G$.
In terms of Laurent modes this is equivalent to
\begin{equation}
  \comm{J^a_n}{J^b_m} = -\frac{n k_L }{4\pi^2} g^{ab} \delta_{n,-m} + \frac{{f^{ab}}_c}{2\pi}J^c_{n+m}.
\end{equation}
The same equations apply for the antiholomorphic currents \(\bar J^a \), the modes \(\bar J^a_n\) and the anomaly coefficient \(k_R\).

In analogy to what we had seen for the energy-momentum tensor, the boundary condition for the general Noether current is that the current perpendicular to the boundary vanishes.
In formulas:
\begin{equation}
  J^a(z)=\bar J^a(\bar{z})
\end{equation} 
and in terms of Laurent modes, this means that a boundary state $\ket{B}$ must satisfy the condition
\begin{equation}
  \pqty{J^a_n + \bar J^a_{-n}}\ket{B}=0 .
\end{equation}
By similar manipulations as the ones above, it is easy to see that this implies
\begin{equation}
  \pqty{ \comm{J^a_n}{J^a_{-n}}- \comm{ \bar J^a_n}{\bar J^a_{-n}} } \ket{B}=0
\end{equation}
which is satisfied for a non-trivial boundary \(\ket{B}\) if and only if the anomaly coefficients agree
\begin{equation}
  k_L = k_R .
\end{equation}

\heading{Intuitive understanding using quantum information theory}

Quantum information theory provides an intuitive argument for
the fact that boundaries are inconsistent with a nonzero gravitational
anomaly.
Suppose $C_L > C_R$.
Consider a packet of  purely left-moving excitation hitting the wall.
At sufficiently high energies $E$, the number of available states goes like $\Delta E \exp[2\pi  \sqrt{\frac{C_L E}{6}}] $ where $\Delta E$ is the energy resolution.
After a time, all states will have reflected from the wall, and turned into right-moving states.
But the number of possible right-moving states is $\Delta E \exp[ 2\pi  \sqrt{\frac{C_R E}{6}}]$.
If $C_L > C_R$, then information cannot be preserved in the sector of sufficiently high energies~\cite{Cardy:1986ie}, which would be inconsistent with unitarity.
The same is true for $C_L < C_R$.
The number of purely left- and right-moving states can only be equal, allowing unitarity to hold in reflection processes involving a single boundary only if $C_L$ and $C_R$ are equal.  In sec. \rr{sec:factorization-implies-boundary} we shall contextualize this quantum information-theoretic
way of understanding
the boundary no-go theorem, by drawing a general logical connection
between boundary conditions and tensor factorizations, deriving a
boundary condition from tensor factorization data in a particular scaling limit.

\subsection{A generalized Nielsen--Ninomiya theorem}
\label{sec:generalized-NN-theorem}

It has been known for a long time that there is a close relationship between anomalies and the (non) existence of a lattice regulator for a given \ac{qft}.
Nielsen and Ninomiya proved various no-go theorems for chiral fermions~\cite{Nielsen:1980rz,Nielsen:1981xu,Nielsen:1981hk} and in particular they showed that it is impossible to get a free relativistic fermionic theory with a different number of left-moving and right-moving modes \(n_L \neq n_R\) as the continuum limit of a lattice theory.

The two statements that we have proved in the previous section, when taken together, give a generalization of this theorem.
\begin{statement}
  \label{lemma:anomaly-implies-no-lattice}
  A two-dimensional \acl{cft} with non-vanishing gravitational anomaly does not admit a lattice regulator.
\end{statement}
The proof is immediate. By statement~\ref{lemma:lattice-implies-boundary}, if \(\cC\) admits a lattice regulator, then it admits a conformal \ac{bc}, but then, by statement~\ref{lemma:boundary-implies-no-anomaly}, the \ac{bc} is non-trivial only if \(C_L = C_R\). Statement~\ref{lemma:anomaly-implies-no-lattice} is the logically equivalent contrapositive. \qed

\bigskip

For a free massless fermion theory, the central charges are just half the number of left or right-moving fermion modes \( c_{L,R} = n_{L,R} /2  \), so statement~\ref{lemma:anomaly-implies-no-lattice} implies the Nielsen--Ninomiya theorem for the pure
gravitational anomaly in $D=2$.
Our proof is more general, though, since it applies to any \emph{interacting} \ac{cft} in two dimensions, even arbitrarily strongly
coupled or nonlagrangian two-dimensional \acp{cft}.

\bigskip

In fact, while conformal invariance is used in the proofs of the previous sections, the final result in statement~\ref{sec:generalized-NN-theorem} can be formulated without referring to conformal invariance, which suggests a further generalization of the Nielsen--Ninomiya theorem.
The idea is that if a generic \ac{qft} \(\cQ\) at energy \(\Lambda_Q\) admits a lattice regulator at energy \(\Lambda_{\textsc{uv}}\), then it represents a point in an \ac{rg} flow that joins this lattice theory with a \ac{cft} in the \ac{ir} at energy \(\LCFT\) where we can use our previous results.

Take a two-dimensional relativistic unitary \ac{qft} \(\cQ\).
The monotonicity of \(C_{L,R}\) implies that the \ac{rg} flow must eventually reach a (possibly trivial) \ac{cft} \(\cC\) at energy \(\LCFT\).
Now, anomaly matching~\cite{tHooft:1979bh} implies that the anomaly coefficient is constant under \ac{rg} flow in any even dimension.
This means that the difference, \emph{e.g.}, the gravitational anomaly, \(C_L - C_R\) is constant, and \(\cC\) has the same gravitational anomaly as \(\cQ\).
Now, if \(\cQ\) has a lattice regulator in the \ac{uv}, by construction so does \(\cC\) and then by statement~\ref{lemma:anomaly-implies-no-lattice} it must have vanishing gravitational anomaly.
It follows that also \(\cQ\) must have \(C_L = C_R\).
All in all we have found a generalization of the Nielsen--Ninomiya theorem to any two-dimensional \ac{qft}.
\begin{statement}
  \label{lemma:anomaly-implies-no-lattice-always}
  A two-dimensional \acl{qft} with non-vanishing gravitational anomaly, does not admit a local lattice regulator with finite entropy per site. 
\end{statement}

\section{Gravitational anomaly obstructs tensor factorization in two-dimensional QFTs}
\label{sec:quantum-info-refinement}

\subsection{Tensor factorization implies existence of boundary CFT}
\label{sec:factorization-implies-boundary}

The generalization of the Nielsen--Ninomiya theorem is an interesting result in its own right, but it is also closely related to some information-theoretic ideas.
The link is given by the fact that the existence of a lattice regulator implies the tensor factorization of the Hilbert space, which is a prerequisite for the very definition of entanglement.
To make the connection precise, we need to refine statement~\ref{lemma:lattice-implies-boundary} about lattice \ac{uv} regulators and the existence of a \ac{bcft}.
In this section we will use the ``renormalized tensor product'' discussed in the Introduction and
indicated by \(\tot\), to allow for local modifications of the naive tensor-product structure at the entangling surface~\cite{Casini:2013rba,Harlow:2015lma}.
\begin{statement}
  \label{lemma:factorization-implies-boundary}
  Let \(\Sigma = A \cup B\) be the spatial slice of a two-dimensional \ac{cft}.  Suppose
  the total Hilbert space factorizes (either exactly or in the generalized sense $\tot$) as the tensor product of two Hilbert spaces associated to the parts \(\Hilb_\Sigma = \Hilb_A \tot \Hilb_B\), %
 in a way that respects the locality of the operator algebra
 in the sense made precise in the Introduction. %
 Then there exists a unitary, energy-preserving
  boundary condition for the \ac{cft} \(\cC^A\) living on each of the two regions, which has finite boundary entropy in the sense~\cite{Affleck:1991tk} of Affleck and Ludwig. 
\end{statement}
In other words, if the Hilbert space is factorizable, starting from the \ac{cft} Hamiltonian \(H_\Sigma \) we can define a Hamiltonian  \(H_A\) that is locally the same as \(H_\Sigma \) (\emph{i.e.}, locally implements the same operator equations of motion) on the region \(A\) and represents a \ac{cft} with boundary.
It is important to emphasize that, while every separable Hilbert space can always be decomposed in the form of a tensor product, we are imposing a stronger condition, \emph{i.e.} that the tensor factors can be thought of as representing Hilbert spaces living in complementary regions of the spatial slice.
This requires the existence of a basis of completely factorized states \(\Hilb = \{\cc \ket{m,n}_\Sigma = \ket{m}_A \otimes \ket{n}_B \}\) and that on such states there is a vanishing correlation between local operators in \(A\) and local operators in \(B\).

By ``tensor factorization that respects the locality of the operator algebra'' we mean something rather specific: We mean a map from the full Hilbert space
to the tensor product %
Hilbert space, such that the factorization map times its conjugate,
is given by a product of approximately-local operators at each component of the entangling surface.  Intuitively\footnote{This intuition is fine, 
but the result is actually somewhat stronger than this, because it excludes \emph{e.g.} the construction of~\cite{Holzhey:1994we}, which
is nonlocal on the scale of the subregion itself.  See the discussion
in the Appendix.} this means that the
map leaves unaffected
the equations of motion for the fields in the interior of regions $A$ and $B$. We make this concept slightly more precise in Appendix \ref{LocalityCriterionPreciseDefinition}.

The proof goes similar in spirit to that of Statement \ref{lemma:lattice-implies-boundary}.
In short, we would like to integrate out the degrees of freedom on $B$ by tracing all the degrees of freedom in $B$ out from the full renormalized Hamiltonian $H$, and construct the Hamiltonian $H_A$ define on $A$, with some boundary condition.
For simplicity, we take $A$ and $B$ to be right and left half of the spatial slice, each defined by $x\leqslant 0$ and $x\geqslant 0$.
Intuitively, such a procedure seems easily conducted as follows,
\begin{equation}
\label{meaningless}
H_A^{\rm trial}\equiv \frac{\Tr_B\left[H\right]}{\Tr_B[\mathbb{1}]}
\quad \text{($\leftarrow$Meaningless definition containing divergences.)},
\end{equation} 
but this expression suffers from divergences both from the numerator and the denominator, for continuum theories.

We could then use some other definitions by suitably regularizing by a short distance scale $\epsilon$ and taking a limit $\epsilon\to 0$.
We denote the regularized half-space Hamiltonian as $H_A^{\epsilon}$.
Not so intuitive but a convenient definition for $H_A^{[\epsilon]}$ we adopt is the following,
\begin{equation}
H_A^{[\epsilon]}\equiv -\frac{1}{\epsilon}\log \left(\Tr_B\left[e^{-\epsilon H}\right]\right)
-(\text{divergent vacuum energy}).
\end{equation}
This actually has a convenient name in quantum information theory, which is the ``high-temperature modular Hamiltonian''.

The more physically intuitive definition $\widetilde{H_A^{[\epsilon]}}$ can be given by a slight modification from replacing $1/\epsilon$ with $d/d\epsilon$, so that
\begin{equation}
\widetilde{H_A^{[\epsilon]}}\equiv
-\frac{d}{d\epsilon}\log \left(\Tr_B\left[e^{-\epsilon H}\right]\right)
=
\frac{\Tr_B\left[He^{-\epsilon H}\right]}{\Tr_B\left[e^{-\epsilon H}\right]},
\end{equation}
where one is able to see that $\epsilon$ works as the ultraviolet cut-off introduced to~\eqref{meaningless}.
Note that both ${H_A^{[\epsilon]}}$ and $\widetilde{H_A^{[\epsilon]}}$ have the same locality property to all orders in $\epsilon$, because the expressions match at each order in $\epsilon$ modulo numerical coefficients.
Especially, they match if one takes the limit $\epsilon\to 0$.

Now let us analyse the locality property of the high-temperature modular Hamiltonian, ${H_A^{[\epsilon]}}$.
As we have already stated in the introduction, we specifically like to think about partial traces which only modifies the density matrix (almost-)locally.
This means the insertion of $\mathcal{F}[\mathcal{M}]$ (whose non-locality only has size of order $\epsilon_{\rm f}<\epsilon$) at the entangling surface will never modify the density matrix $e^{-\epsilon H}$ at distance $\epsilon$ away from the entangling surface.
This guarantees the same commutation relation of $e^{-\epsilon H_A^{[\epsilon]}}$ with the original $e^{-\epsilon H}$, so that we have
\begin{equation}
e^{-\epsilon H_A^{[\epsilon]}} \mathcal{O}(0, x)
=\mathcal{O}(\epsilon,x)e^{-\epsilon H_A^{[\epsilon]}}
\quad (x>\epsilon).
\end{equation} 
And by taking a derivative in $\epsilon$, we get
\begin{equation}
\left[H_A^{[\epsilon]},\mathcal{O}(0,x)\right]
=\partial_t{\mathcal{O}}(0,x)
\end{equation}
and hence we see that $H_A^{[\epsilon]}$ can be represented as a sum of local operators $H_A^{[\epsilon]}(x)$, where
\begin{equation}
H_A^{[\epsilon]}(x)=H(x)\quad (x>\epsilon).
\end{equation}

Now, it is particularly easy if one is in the bulk of $A$, \emph{i.e.}, in $x>0$ and take the high-temperature limit $\epsilon\to 0$.
The high-temperature modular Hamiltonian simply becomes
\begin{equation}
H_A(x)=H(x)\quad (x>0).
\end{equation}
Things are different on the boundary, $x=0$, however.
In general, we expect relevant boundary operators with coefficients that scale as \(\epsilon^{\Delta - 1}\) (\emph{e.g.} there will be a boundary cosmological constant scaling as \(\epsilon^{-1}\)).
Such operators are also renormalized in the usual Wilsonian sense as we take $\epsilon\to 0$.
Again the Friedan--Konechny hypothesis ensures that such a procedure terminates at some point and gives us a finite result.
In other words, a boundary \ac{rg} flow for a \ac{cft} will lead the theory on the half-space to a \ac{bcft} with finite boundary entropy.
This concludes the proof of statement~\ref{lemma:factorization-implies-boundary}. \qed

\subsection{A no-go theorem for tensor factorizations}
\label{sec:quantum-information}

Combining statement~\ref{lemma:factorization-implies-boundary} with statement~\ref{lemma:boundary-implies-no-anomaly} we can generalize statement~\ref{lemma:anomaly-implies-no-lattice} and prove that a gravitational anomalous theory cannot be factorized (in the sense that we have defined above).

\begin{statement}
  \label{lemma:anomaly-implies-no-factorization}
  A two-dimensional gravitationally anomalous theory defined on \(\Sigma = A \cup B\) does not admit a Hilbert space factorization \(\Hilb_\Sigma = \Hilb_A \tot \Hilb_B\),
  even in any sense generalized by local short-distance modifications at the entangling surface.
\end{statement}
It follows that there is no possible consistent definition of entanglement respecting the locality of the operator algebra.

\section{Obstruction to entanglement for six-dimensional theories}
\label{sec:six-dimensions} 

Up to this point we have only discussed the factorization and entanglement for anomalous two-dimensional field theories.
In fact, via dimensional reduction, we can deduce consequences also in higher dimensions.
In this section, we will concentrate on the case of six-dimensional theories. 
\begin{statement}
  \label{lemma:six-dimensional-entanglement}
  A six-dimensional theory with non-vanishing factorized anomaly coefficient cannot have entanglement across a surface with non-zero signature class.
\end{statement}
This implies in particular that the \((2,0)\) superconformal theory in six dimensions (whose factorized anomaly coefficient is proportional to the number of tensor multiplets) does not admit a tensor factorization across an entangling surface such as \(\mathbb{CP}^2\).

To prove our statement, consider the six-dimensional (purely gravitational) anomaly polynomial for a theory living on a six-manifold \(X\)
\begin{equation}
  I_8(X) = k_{\text{irr}} \Tr(R^4) + k_{\text{fact}} \Tr(R^2)^2
\end{equation}
where \(k_{\text{in}}\) and \(k_{\text{fact}}\) are respectively the irreducible and factorized anomaly coefficients that depend on the matter content of the theory (see~\cite{Avramis:2006nb} and references therein for explicit examples).
Equivalently, \(I_8\) can be rewritten in terms of the first and second Pontryagin classes on the tangent bundle \(TX\):
\begin{equation}
\label{AlphaBetaDefs}
  I_8 = \frac{1}{4!} \pqty{\delta \,  p_2(TX) + \gamma \, p_1(TX)^2} ,
\end{equation}
where the coefficients \(\delta\) and \(\gamma\) are related to \(k_{\text{irr}}\) and \(k_{\text{fact}}\) by
\begin{align}
\label{KIrrAndKFactDefs}
  k_{\text{irr}} &= - \frac{\delta}{3 \times 2^9 \pi^2} & k_{\text{fact}} &= \frac{\delta + 2 \gamma}{3 \times 2^{10} \pi^4} .
\end{align}

The idea is to compactify \(X\) on a four-manifold \(K\) to a two-dimensional surface \(\Sigma\) and write the anomaly polynomial \(I_4(\Sigma)\) integrating \(I_8(X)\) over \(K\).

The Pontryagin classes are expressed in terms of the Chern roots \(\set{\pm l_i}\) as
\begin{align}
  p_1(TX) &= \sum_i l_i^2 & p_2(TX) = \sum_{i < j} l_i^2 l_j^2 . 
\end{align}
Let \(\set{\pm t}\) and \(\set{\pm \lambda_{1,2}}\) be respectively the Chern roots on \(\Sigma\) and on \(K\), then the anomaly polynomial \(I_8(X)\) is
\begin{equation}
  I_8(X) = \frac{1}{4!} \pqty{\gamma t^4 + \pqty{\delta + 2 \gamma} \pqty{ \lambda_1^2 + \lambda_2^2} t^2 + \gamma \pqty{\lambda_1^2 + \lambda_2^2}^2 + \delta \lambda_1^2 \lambda_2^2}. 
\end{equation}
It is clear that if we integrate \(I_8(X)\) over \(K\) to get \(I_4(\Sigma)\), only the second term gives a non-vanishing contribution.
Then
\begin{equation}
  I_4(\Sigma) = \int_K I_8(X) = \frac{\delta + 2 \gamma}{4!} t^2 \int_K \pqty{\lambda_1^2 + \lambda_2^2}
\end{equation}
where we recognize the first Pontryagin classes on \(\Sigma\) and \(K\):
\begin{equation}
  I_4(\Sigma) = \frac{\delta + 2 \gamma}{4!} p_1(T\Sigma) \int_K p_1(TK) =  \frac{\delta + 2 \gamma}{8} p_1(TX) \tau(K) , 
\end{equation}
where \(\tau(k)\) is the signature of \(K\).
On the other hand, in general \(I_4(\Sigma)\) is related to the central charges by
\begin{equation}
  I_4(\Sigma) = \frac{C_L - C_R}{24} p_1(T\Sigma) ,
\end{equation}
hence
\label{CentralChargeDiff}
\begin{equation}
  C_L - C_R = 3 \pqty{\delta + 2 \gamma} \tau(K) = \pqty{96 \pi^2}^2 \cc  k_{\text{fact}} \tau(K).
\end{equation}
Using statement~\ref{lemma:anomaly-implies-no-factorization} we see that unless \(\tau(K) = 0\) no factorization is possible across an entangling surface \(K\).\qed

The factorized-gravitational-anomaly coefficient $\pqty{\delta + 2 \gamma}$ is generally nonzero in interesting chiral theories,
and even in chiral theories that are boring because they're free.  In theories with at least the minimal supersymmetry $(1,0)$ 
in six dimensions, an up-to-date set of
known values of $\gamma$ and $\delta$ for various supersymmetric multiplets and field theories
are given  \emph{e.g.} in reference~\cite{Cordova:2015fha} and reproduced in Table~\ref{tab:known-anomalies}%
\def\ssc{\hskip-.5in}
\def\sssc{}

\def\ssc{{\color{White} AAA}}
\def\wwa{{\color{White} A}}
\def\wwd{{\color{White} .}}

\begin{table}[h]
  \hspace{-.5cm}
  { \def\arraystretch{1.5}
    \begin{tabular}{lR{5.5em}R{5.5em}R{5em}R{1em}}
      \toprule
      Theory                                            & $\gamma$                         & $ \delta$                   & \multicolumn{2}{r}{$\pqty{C_L - C_R}/\tau(K)$}                         \\
      \midrule
      scalar, gauge field                               & \(0\)                            & \(0\)                       & \(0\)                 & $\wwd$                         \\
      Weyl$\ll +$ fermion                               & \(\frac{7}{240}\)                & \(-\frac{1}{60}\)           & \( \frac{1}{8}\)                             \\
      Weyl$\ll -$ fermion                               & $-{7 \over {240}}$               & ${1 \over 60}$              & $-{1\over 8}$                            \\
      Selfdual$\ll {++}$ tensor field                   & ${16} \over 240$                 & $-{{28} \over 60}$          & $-1$                                    \\
      Selfdual$\ll {--}$ tensor field                   & $-{{16} \over {240}}$            & ${{28} \over 60}$           & $1 $                                           \\
      \midrule
      \((1,0)\)~vector multiplet                        & $-{7 \over 240}$                 & $1 \over 60$                & $-{1\over 8} $                                \\
      hypermultiplet                                    & $7 \over 240$                    & $-{1 \over 60}$             & ${1\over 8} $                          \\
      \((1,0)\) tensor multiplet                        & $23 \over 240$                   & $-{29 \over 60}$            & $-{7\over 8}$                                 \\
      \((2,0)\) tensor multiplet                        & ${1\over 8}\cc $                 & $-\hh\cc  $                 & $-{3\over 4}  $                            \\
      \midrule
      hypermultiplet in representation \(R\)            & $  {7\over{240}}\abs{R}$         & $  - {1\over{60}}\abs{R} $  & $   {1\over 8} \abs{R} $                  \\
      \((1,0)\)~vector multiplet with group \(G\)       & $-{{7} \over 240} \abs{G}$       & $\frac{1}{60} \abs{G}$      & $-\frac{1}{8} {\abs{G}}$   \\
      $(1,1)$~vector multiplet with group \(G\)         & \(0 \)                           & \(0\)                       & \(0  \)                                  \\
    $(2,0)$ nonabelian tensor of rank $N$               & ${1\over 8}\cc N$                & $-\hh\cc N$                 & $-{3\over 4}\cc  N $                     \\
      \bottomrule
    \end{tabular}
  }
  \caption{Pure gravitational anomaly coefficients as normalized in~\cite{Cordova:2015fha}. Here $\abs{G} = \dim(G)$ and $\abs{R}$ is the dimension of
    $R$ as a complex representation.  The orientation conventions
    are such that the supercharge itself has chirality $+$, and the orientation convention for the signature $\t$ is that
    $\t(K3) = -16$ for a \(K3\) surface.}
  \label{tab:known-anomalies}
\end{table}

We have listed the values 
corresponding to an untwisted compactification, that is, one where the scalars have only their usual conformal coupling
to the Ricci scalar, vectors and tensors couple only to the metric and Riemannian connection, and fermions couple only to the spin connection on the tangent bundle.  These values for the effective anomaly coefficient $C\ll L - C\ll R$ applies to any compactification manifold $K$ that is admissible for a given theory in $D=6$.  But we remind the reader that not every compactification $K$
is admissible for every field content.  Untwisted compactifications for theories containing spinors are only
possible if $K$ is a spin manifold, and so in particular supersymmetric theories do not have untwisted compactifications
on $K$ unless it is spin.  The simplest spin manifolds in four dimensions are $T\uu 4$ and the complex $K3$ surface,
which has signature $-16$.  It is still possible to compactify theories with spinors on a four-manifold that is only spin$\ll{\setC}$ rather
than truly spin, but this involves coupling the fermions to a background $U(1)$ connection, which shifts the value
of the anomaly, relative to the values listed in the table, by an amount proportional to a linear combination
of mixed gauge-gravitational and pure gauge anomalies for the $6D$ theory.  We do not give these contributions to $C\ll L - C\ll R$
in the table above, but see~\cite{WatanabeThesis:2015} for an analysis of mixed anomalies and constraints on boundary
conditions and tensor factorizations in $D = 4,6$ from mixed anomalies.

There are likely further interesting implications of the anomalous obstruction to local tensor factorizations for six dimensional
theories.  Before closing the section we shall also mention one more implication for quantum field theory, namely
for the intrinsically strongly coupled chiral $(2,0)$-superconformal
theory in six dimensions, which contains rank-two tensor gauge fields with self-dual three-form field
strengths.  Various interesting proposals have been made (see for instance~\cite{Lipstein:2014vca}
for a recent example)
 for possible lattice realizations of such theories.  We note here
that the no-go theorem for six-dimensional theories with nonvanishing $\d + 2\g$-coefficient, implies the impossibility
of a lattice realization of such theories in the strict sense.  The obstruction is related not directly to the nonabelian nature of
the two-form dynamics, but to their chirality: Indeed the anomaly is nonvanishing for even a free Abelian self-dual two-form
theory, the impossibility of a lattice theory flowing to a chiral two-form is visible even at that level.   Of course the no-go
theorem still allows the possibility of modified constructions, such as a lattice theory flowing to two parity-mirror copies
of the strongly interacting $(2,0)$ theory.  However, as discussed elsewhere, such a construction still would not produce
a meaningful notion of reduced density matrices for local Hilbert spaces, that correctly capture the correlations
in the state.

\section{Physical meaning of nonfactorization in QFT}
\label{sec:discussion}

\subsection{Anomalous irreducibility of long-distance correlations}\label{IrreducibleCorrelations}

We have shown that the Hilbert space of a quantum field theory with $C\ll L \neq C\ll R$ can never factorize into 
a tensor product.  In this section we wish to emphasize that this is a physically meaningful, robust obstruction
to factorization, in particular a property of correlation functions even very far from the entangling surface.  It is \rwa{not} 
an artifact of the short-distance details of how the \ac{qft} is regulated or renormalized, nor of how the tensor factorization is defined
in the ultraviolet.  

In particular, we contrast the obstruction discussed in the present paper, with the
phenomenon of nonfactorization discussed in~\cite{Casini:2013rba}, which can exist in any dimension and is not associated
with any anomaly.  The type of obstruction studied in~\cite{Casini:2013rba} and subsequent literature is associated with short-distance
behavior near the entangling surface.  By adding boundary degrees of freedom, imposing boundary
constraints, and modding out by boundary gauge symmetries, we can in examples such as~\cite{Casini:2013rba} factorize
the Hilbert space unproblematically.

This is emphatically \rwa{not} the case with the obstruction to tensor factorization associated with the gravitational anomaly,
which is a long-distance phenomenon, robust against any deformation or modification that leaves the theory unchanged
in the infrared. One striking consequence is that a gravitationally anomalous theory does \rwa{not} contain (pure) quantum states with
vanishing correlations between two disjoint regions $A$ and $B$.  Of course, even in an anomaly-free theory with nonzero
central charge, 
the vacuum has long-range correlations as does any finite-energy state; 
there is always an energy cost to disentangling the two halves of the real line.  But the energy cost is finite: In anomaly-free theories, there is an upper bound to the amount of energy needed to suppress all quantum correlations between disjoint regions by a given amount.  Specifically, there exists%
~\cite{WatanabeThesis:2015} an exponent $\d\ll 1\uprm{max} > 0$ such that correlations between the two halves of the real line can be suppressed by an
amount $(xE)\uu{-2\d\ll 1\uprm{max}}$ at points separated from the boundary by a distance $x$, if one has an available energy $E$. 

 There is no such exponent in a theory with $C\ll L \neq C\ll R$.  If even one state existed with uniformly suppressed
 correlations, then one could act with local operators on the two sides separately, to build a basis of the full Hilbert
 space with vanishing correlations in each basis state.  But then, such a basis would be a construction of a tensor factorization
 of the Hilbert space, which we have shown cannot exist for theories with $C\ll L \neq C\ll R$.  Thus, correlations can \rwa{never} be
 suppressed parametrically relative to the vacuum correlations in any state in a gravitationally anomalous \ac{cft}, however high
 the energy of the state.\footnote{This statement applies to pure states only; in mixed states such as a finite temperature thermal ensemble,
 correlations are uniformly suppressed as usual on scales longer than the inverse temperature.  But this is attributable to cancellations
 due to phases in correlation functions in energy eigencomponents; the magnitudes of correlations in each energy eigencomponent are comparable
 to those in the vacuum state.}
 
 Perhaps not surprisingly, the exponent $\d\ll 1\uprm{max}$
 is associated with the existence of a boundary condition: $\d\ll 1$ is the maximum
 energy gap $E\ll 1\uprm{interval} - E\ll 0\uprm{interval}$ on the interval of length $\pi$, for any unitary, energy-conserving
 boundary condition in the \ac{cft} with finite boundary entropy.  Results on the open-closed modular bootstrap
~\cite{Friedan:2012jk} has established an upper limit on $\d\ll 1\uprm{max}$ in various \acp{cft}, including a very precise bound in the famous
 $C\ll L = C\ll R =24$ monster \ac{cft}, shown to be saturated by a particular boundary condition.  It is
 interesting that the problem of establishing an upper limit on $\d\ll 1$ among all possible boundary conditions, 
 turns out to have fundamental implications for information theory in field theories \rwa{without} boundary.
 
One of the powerful illustrations of such a principle is the following fact:
In party-invariant theories, there always exists the maximal finite energy gap on the interval.
This is dictated by the observation that
there always exists a unitary boundary condition with finite boundary entropy for such systems, by taking a parity-symmetric boundary condition.

\heading{Example: Free fermions}

To illustrate the point above, we consider a free fermion system.  First, consider the case of a massless Majorana fermion $\psi$ in $1+1$
dimensions.  This system has $C\ll L = C\ll R = \hh$, and no gravitational anomaly, although it does have anomalies associated
with the chiral discrete symmetry $\psi\ll L \to - \psi\ll L, \cc \psi\ll R \to + \psi\ll R$.  Nonetheless, the absence of
a pure gravitational anomaly suggests it may be possible to construct (pure) quantum states on $\IR$ in which the fermion
fields at $x < 0$ are approximately uncorrelated with the fermion fields in $x > 0$, to an approximation that improves
as a power law of the total energy budget available
to create the state.

Suitable states can be in principle described directly as squeezed states
in the Fock space of $\psi$, but it is simpler to characterize
the states in an equivalent but more local way, as ground states of a modified local Hamiltonian.  If our conformal, translationally invariant
Hamiltonian for the field is $H\uu {(0)}\ll{\psi}$, then we can define the state $\kket{{\rm barr},~\m}$ as the ground state
of the modified Hamiltonian
\begin{equation}
H\pr\ll{\psi} \equiv H\uu {(0)}\ll{\psi} + H\uu{{\rm barr}}\ll\psi\ ,
\end{equation}
where
\begin{align}
H\uu{{\rm barr}}\ll\psi &\equiv \int \dd{x}  {{\cal H}}\uu{{\rm barr}}\ll\psi\ , \\
{{\cal H}}\uu{{\rm barr}}\ll\psi &\equiv f\ll\m(x) \cc \psi\ll L(x) \psi\ll R(x)\ ,
\end{align}
is a localized mass term for the Majorana fermion, whose profile $f\ll\m(x)$ has a height and width governed by the energy scale $\m$.
The details of $f\ll\m(x)$ are unimportant; we choose it so as to create a potential barrier for the fermion through which it
cannot easily propagate from one half of the real line to the other. 
A Gaussian or simple step function of height $\m$ and width $\m\uu{-1}$ is sufficient to do this.

Regardless of the details of the function $f\ll\m(x)$, the expectation value of the unmodified Hamiltonian $H\uu{(0)}\ll\psi$
in the state $\kket{{\rm barr},~\m}$, goes as $\m$ with a positive coefficient
\begin{align}
  \ev{E}_\mu \equiv \ev{H_\psi^{(0)}}{{\rm barr},~\m} =  K\ll E \cdot \m\ , && K_E > 0
\end{align}
 for some constant $K\ll E$ independent of $\m$ but depending on the shape of the barrier profile $f$,
as dictated by dimensional analysis and the fact that $H\ll\psi\upp 0$ is bounded below (at zero).
For any reasonable choice of barrier profile $f\ll\m(x)$, the correlation functions between fields on the two separate halves
of the real line, can be suppressed by an amount proportional to $\m\uu{-1}$, relative to the size of the correlation in the
ground state.  That is, if
\begin{align}
  \abs{ \ev{\psi(x) \psi(-x)}{{\rm barr},~\m}} \lesssim \frac{K\ll{\psi\sqd}}{\mu x} \cdot 
\frac{1}{\mu x}\ , && x \muchgreaterthan \m\uu{-1}\ ,
\end{align}
 for some constant $K\ll{\psi\sqd}$ independent of $\m$ and
 depending on the shape $f$ of the barrier profile, whereas the vacuum two-point function
 of $\psi$ goes as $|x|\uu{-1}$.  A ground state of a quadratic Hamiltonian always has $n$-point functions
 given by a sum of 
 Wick contractions, so any two-point function in
 the barrier state of any operator in the theory, is also suppressed by a factor of $|\m x|\uu{-1}$ at least, relative to the value of that same two-point function in the ground state.
 
 Thus, we find we can suppress correlations of
 the two sides of the real line with a strength of $\m\uu{-1}\cc
 E\lrm{infrared}$, if we are given an available energy
 budget of $\m$. In terms of the recipe in
\cite{WatanabeThesis:2015} for discorrelating sides
of the real line, this suppression $\m\uu{-2\d\ll 1} = \m\uu{-1}$ exponent corresponds
 to the first-excited state on the interval with
a specified boundary condition.

Specifically, if we consider the interval with NS boundary
conditions for the fermion,
$\psi\ll L = +\psi\ll R$ at $\s = 0$ and $\psi\ll L = - \psi\ll R$ at $\s = \pi$, then the first excited energy is
\begin{equation}
  \d\ll 1 = E\ll 1\uprm{interval} - E\ll 0\uprm{interval} = +\hh\ .
\end{equation}
This illustrates the discorrelation principle~\cite{WatanabeThesis:2015}, in the concrete example of
a nonchiral fermion theory.
 
 The case of a chiral theory is different.
 For a Majorana-Weyl fermion, the central charges
 are $C\ll L = +\hh, C\ll R = 0$, and there is a gravitational
 anomaly. We have already proven
 abstractly that we cannot arbitrarily strongly
 suppress correlation functions in a pure state between two
 halves of a real line in this case.  But it is easy
 to see concretely that we cannot do so in any
 squeezed state with finite expectation value for
 the energy.
 
 The tools available in the nonchiral fermion
 theory to disentangle the two sides
 of the real line, do not exist in the chiral fermion
 theory.  The only local operators available have nonzero
 spin, and in particular there are no relevant operators with
 Bose statistics.  In a theory with a complex rather than real Weyl fermion, one can modify the Hamiltonian by 
 adding a term
 \begin{equation}
   \label{ChiralCurrentModifiedHamiltonian}
   \begin{aligned}
H\pr &\equiv H\uu {(0)} + H\uu{{\rm barr}}\ll\psi\ ,\\
H\uu{{\rm barr}} &\equiv \int \dd{x} {{\cal H}}\uu{{\rm barr}}\ , \\
{{\cal H}}\uu{{\rm barr}} &\equiv f\ll\m(x) \cc \psi\dag\ll L(x) \psi\ll L(x)\ .
\end{aligned}
 \end{equation}
This term also grows in importance in the infrared,
but its effect is qualitatively different
from that of the localized Majorana mass term in the
case of the nonchiral fermion.  For a complex
Weyl fermion, the current is chiral, and so has
only one independent component, and
so the charge density equals the spacelike component of the current density, $J\uu 0 = J\uu 1$.  Adding the term
\rr{ChiralCurrentModifiedHamiltonian} to the Hamiltonian
does not reduce the correlation between the two sides; in fact, on the infinite line, it has no physical
effect at all and can be absorbed into a time-independent redefinition of the field $\psi$. 

The current $J\ll L = \psi\dag\ll L \psi\ll L$ is the only non-identity operator in the theory with Bose statistics and dimension
less than 2.  Deforming the Hamiltonian near the origin by adding operators of dimension 2 or higher, could only be
expected to have a constant or decreasing effect at long distances, and so would not suppress correlation functions.  Similar comments apply if we have three or more Majorana-Weyl fermions of the same chirality.

If it were not for the anomaly there would be the possibility that a suitable choice of modification with a non-infinitesimal
coefficient, could still run to an interface theory with a nontrivial fixed-point behavior that could screen correlations between
the two sides, but in a simple free fermion theory this is unlikely \it ex ante \rm and in a fermion theory with $C\ll L \neq C\ll R$ we have shown it is impossible on general grounds.  
 
 \subsection{General construction of approximately-factorized states in theories admitting boundary conditions}\label{ApproxFacStateConstruction}
 
 The construction of increasingly discorrelated states as we raise the energy, is a specific example of a more general construction,
first appearing in~\cite{WatanabeThesis:2015}.  We now describe the construction in the more general case, to illustrate that the tight connection between the existence of boundary conditions, and the existence of factorized states, can be made precise at finite energy.

\heading{The (anti-)slit geometry}
Of course there can be no perfectly factorized states in a continuum field theory, with finite energy; however we wish to demonstrate
that, in a theory admitting a boundary condition, one can always construct states that are factorized between regions to any desired approximation,
by adding a sufficiently large amount of energy, of an order of magnitude we characterize by the generic energy scale $\L$. 

\begin{figure}[htbp]
 \begin{center}
 \begin{overpic}[width=130mm]{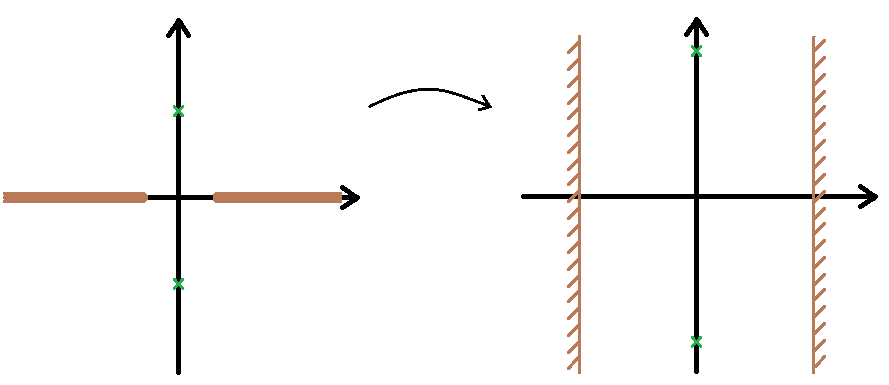}
 \put(36,36){$f(z)=\arcsin(\Lambda z)$}
 \put(25,17){$1/\Lambda$}
 \put(11,17){$-1/\Lambda$}
 \put(22,10){$-i\sigma/2$}
 \put(22,30){$i\sigma/2$}
 \put(42,20){$t$}
 \put(19,42){$x$}
 \put(78,42){$t$}
 \put(100,20){$x$}
 \put(80,36){$i\ell/2$}
 \put(80,4){$-i\ell/2$}
\end{overpic}
 \end{center}
 \caption{The slit geometry can be conformally transformed
 into a strip. The crosses represent the positions at which the
 local operators are located.
 We may take whatever boundary condition as long as it has finite boundary entropy.
}
 \label{slit}
\end{figure}
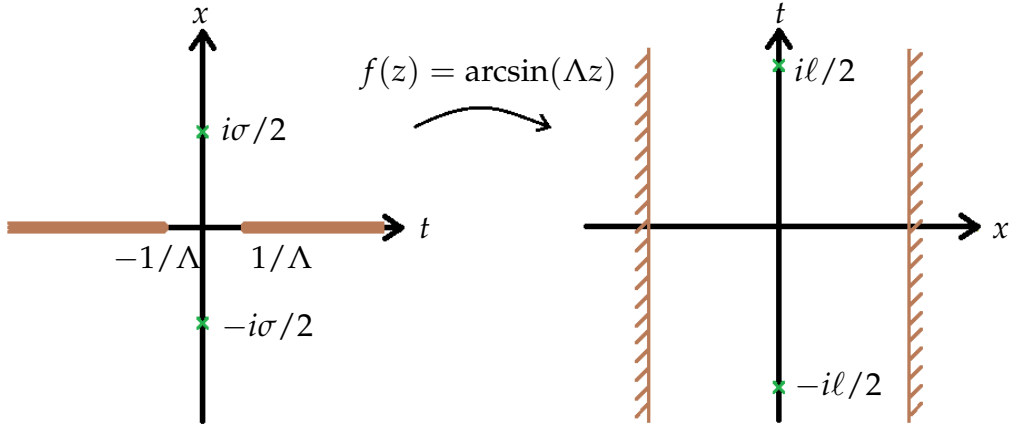

We construct the approximately-factorized state by considering a path integral on the following geometry: We have a boundary at $x=0$
running from $t=-\infty$ to $t=-1/\Lambda$ and 
then from $t=1/\Lambda$ to $t=\infty$,
with no barrier between $t=-1/\Lambda$ and $t=1/\Lambda$
(see Fig. \ref{slit}).
This ``slit'' geometry\footnote{Though referred to as a ``slit'' geometry, 
the terminology is not perfectly accurate, since really there are two
semi-infinite slits in the center of the geometry, with a small gap between
the slits, connecting the two sides.  Though it would be more accurately
described as an ``anti-slit'' geometry, we will stick to the existing
term.} (first appeared in the analysis of local quenches in~\cite{Calabrese:2007mtj}) can be conformally transformed to 
the ordinary open-channel strip (see again Fig. \ref{slit}):
\begin{equation}
z\mapsto w=\arcsin (\Lambda z).
\end{equation}
Specifically, the point $z=i\sigma/2$ 
in the slit geometry is mapped to the point
\begin{align}
  w&=\frac{i\ell}{2}
     =i\log\left(\Lambda\sigma+\sqrt{(\Lambda\sigma)^2+1}\right)
     \sim i\log\left(\Lambda\sigma\right) \quad \text{if \(\Lambda\sigma \muchgreaterthan 1\)}
\label{gggg}
\end{align}

\heading{The vanishing of correlation}

We prove that the correlation between the same operator $\Op$, whose dimension is $\Delta$, located at $x=\sigma$ and $x=-\sigma$ vanishes as $\Lambda\to\infty$.
We compute the connected two-point function in the slit geometry,
\begin{equation}
  \frac{\braket{\Op(-i\sigma)}{\Op(i\sigma)}_c}
  {\braket{\mathrm{vac}}}
\end{equation}
and take the limit $\Lambda\sigma\to\infty$.
We then move to the strip frame. By expanding the bra and the ket in terms of eigenstates of the Hamiltonian in the open-channel strip and considering the Weyl scaling correctly, we get
\begin{equation}
  \frac{\braket{\Op(-i\sigma/2)}{\Op(i\sigma/2)}_c}
  {\braket{\mathrm{vac}}}
  \propto e^{-\ell (E_1-E_0)}\times \sigma^{-2\Delta},
\end{equation}
where $E_1-E_0\equiv \Delta_1$ is the first excitation energy of the system measured from the lowest state energy.
Therefore, by using Eq. (\ref{gggg}), we get
\begin{equation}
  \frac{\braket{\Op(-i\sigma)}{\Op(i\sigma)}}
  {\braket{\mathrm{vac}}}\propto (\Lambda\sigma)^{-2\Delta_1}\times \sigma^{-2\Delta}\to 0
  \label{pls}
\end{equation}
and the two-point function vanishes in the limit of $\Lambda\to\infty$ as promised.
This explicitly shows that almost factorized states give damped
correlation between the two sides power-law in $\Lambda$, the available energy.

\heading{Physical meaning}

Instead of the geometric argument given above, we could also equivalently
consider a sum of local effective operators
connecting the two regions divided by the wall:
\begin{equation}
{\Op}_{\mathrm{hole}} \equiv 
\sum_i c_i \Lambda^{-2 \delta_i} 
{\Op}^{i}_{\text{left}}  {\Op}^{i}_{\text{right}}.
\end{equation}
Then the leading two-point function is 
$(\sigma^{-2\Delta - 2\delta_1}) \Lambda^{-2\delta_1}$, which agrees
at the level of scaling with the above result.
Note that this effective operator of size $1/\Lambda$ is an example of $\mathcal{F}[\mathcal{M}]$ introduced earlier.

\subsection{Holographic theories with large gravitational anomalies}
\label{sec:holography-and-anomaly}

\heading{Known examples of anomalous holographic \ac{cft}}

As mentioned in the introduction, the anomalous obstruction to tensor factorization of the Hilbert space is present even 
in some examples which exist in the holographic regime and have apparently well-behaved, ultraviolet-complete gravity duals.
The case of ${\cal N} = 4$ SYM has no gravitational anomaly, and has lattice versions~\cite{Kaplan:2002wv}, some of which
\cite{Hellerman:2002qa} even have concrete holographic duals whose  (continuum) geometry and flux encodes
the effect of the lattice structure of the boundary theory.  Indeed, pure gravitational anomalies do not even exist
in four dimensions.  However, there are other cases of gauge-gravity duality in two and six dimensions, where the pure
gravitational anomaly coefficient is not only nonzero but large, although not as large as the total central charge,
leaving it in the middle range \rr{HolographicAnomalyHierarchy}.

\bigskip

One explicit example includes the M-theory solutions of the form \(AdS_3 \times S^2 \times CY_3 \) in~\cite{Gauntlett:2000ng} dual to the \((0,4)\) \ac{scft} introduced in~\cite{Maldacena:1997de}.
In this case, the gravitational anomaly has been computed, and it is proportional to $N\ll{\rm M5}\cc \int \ll D \cc c\ll 2(T{D})$, where
$D$ is a divisor (complex codimension-$1$ submanifold) in a Calabi--Yau threefold, and $c\ll 2(TD)$ is the second Chern class of the tangent bundle to $D$.
The difference between
the $c_2(TD)$ here and the signature of $D$ appearing in the anomaly analysis of section~\ref{sec:six-dimensions} is
related to the coupling of the scalars to the connection on the normal bundle of $D \subset CY_3$; this coupling is needed to preserve supersymmetry, and alters the anomaly polynomial.%
\footnote{For anomalies in other \ac{scft} obtained by compactifying M5-branes on supersymmetric cycles, see recent work such as~\cite{Benini:2012cz, Benini:2013cda, Gadde:2013sca}; for the anomalies of the M5-brane theory in a general background see~\cite{Ohmori:2014kda}.  For further development of M-theory solutions dual to more general chiral two-dimensional \ac{scft}  see \emph{e.g.}~\cite{Bah:2015nva}.}
Little is known about the spectrum or \ac{ope} coefficients of the $(0,4)$ theory beyond the anomalies, and some supersymmetric invariants such as the elliptic genus~\cite{Gaiotto:2006wm, Gaiotto:2007cd}.
However, the total central charge $C\ll L + C\ll R$ goes
as $N\ll{M5}\uu 3$ times the triple self-intersection number of the divisor in the ambient $CY_3$, giving
\begin{align}
   C_L + C_R &\propto N_{M5}^3  , &  \abs{C_L - C_R} &\propto N_{M5} \sim (C\ll L + C\ll R)\uu{{1\over 3}}
\end{align}
which satisfies the condition \rr{HolographicAnomalyHierarchy}.

Another example is the D1--D5 system described by the \(AdS_3 \times S^3 \times M_4 \) solutions in type I string theory of~\cite{Oz:1999it}. This string theory is also dual to a  $(0,4)$ superconformal field theory with a large gravitational anomaly.
Here the central charges can be calculated easily at weak IIB string coupling, and the total
central charge is of order $Q\ll 1 Q\ll 5$ while the anomaly equal to $c\ll {\rm L} - c\ll{\rm R} = 12\cc Q\ll 1$.
The charges $Q\ll 1$ and $Q\ll 5$ must be scaled together to infinity to keep the moduli of the theory in the holographic regime.  And so, in this regime $Q\ll 5 \sim Q\ll 1\muchgreaterthan 1$, the total central charge and the anomaly are
\begin{align}
  C\ll L + C\ll R &\propto Q\ll 1\sqd , & \abs{C\ll L - C\ll R} &= 12 Q\ll 1 \sim (C\ll L + C\ll R)\uu{{1\over 2}}\ .
\end{align}
satisfying the double hierarchy \rr{HolographicAnomalyHierarchy}.

\bigskip

\heading{Size of the anomaly never of order $c$ in known holographic \ac{cft}}

It would be interesting to understand the reason for the parametric inequality in \rr{HolographicAnomalyHierarchy},
in particular whether there may be a general limitation on
the size of the gravitational CS term~\cite{Deser:1982vy,Deser:1981wh} in a consistent UV-completion in $D=3$ gravity
in AdS$\ll 3$.  Such a limitation could result from a conformal bootstrap inequality and/or a bulk causality constraint.

The study of $D=3$ gravity with a CS term $\m$ of order $G\ll N\uu{-1}$ has been termed ``topologically massive gravity''
~\cite{Deser:1982vy,Deser:1981wh} and its internal consistency as a quantum gravity theory in AdS$\ll 3$ has been explored via holographic duality~\cite{Li:2008dq,Li:2008yz} with a hypothetical dual \ac{cft} of unknown characterization.  Such a treatment considers gravitational CS terms whose magnitude $\m$ is of the same order as the inverse Newton constant $G\ll{\rm N}\uu{-1}$.  

The consistency of this regime is unclear.   The constraint of unitarity of the bulk effective theory yields an inequality
between $\m$ and $G\ll{\rm N}\uu{-1}$ which reproduces the expected condition
\bbb
\abs{C\ll L - C\ll R} \leq C\ll L + C\ll R
\eee
that follows from positivity of the individual central charges $c\ll{\rm L,R}$ in a unitary \ac{cft}.
At the same time, a gravitational CS coefficient $\m$ that is equal to, or smaller than but of the same order as $G\ll {\rm N}\uu{-1}$,
leads to various
exotic classical and semiclassical behaviors, the quantum mechanical consistency of which has never been fully clear
(see~\cite{Li:2008dq} itself or for instance~\cite{Giribet:2008bw} for early observations
along these lines.)
On the \ac{cft} side, as explained in~\cite{Li:2008dq}, the regime $\m\lesssim G\ll{\rm N}\uu{-1}$ would correspond to a gravitational anomaly
of the same size as the total central charge, 
\begin{equation}%
  \label{LSSTMGRegime}
  \abs{C\ll L - C\ll R}\lesssim C\ll L + C\ll R\ .
\end{equation}

There are no known examples
of \ac{cft} in the regime \rr{LSSTMGRegime}
which are holographically dual to Einstein gravity plus a finite number of light fields.  For instance,
free field theories or other high-order products of low-$c$ conformal theories, can easily be constructed in this regime,
but they do not correspond to the holographic regime describing gravity coupled to field theory, which
requires a gap~\cite{Heemskerk:2009pn} in operator dimensions, 
\begin{equation}%
  \label{PPHSPlusModularHolographicConditions}
  1 \muchlessthan \D\ll 1 \lesssim O(c) \leq {{C\ll L + C\ll R}\over{12}}
\end{equation}
the upper bound being
forced by modular invariance~\cite{Hellerman:2009bu,Friedan:2013cba,Collier:2016cls,Afkhami-Jeddi:2019zci,Hartman:2019pcd}.

 In other words, no known complete quantum theory has been able to attain the macroscopic size \rr{LSSTMGRegime} of the gravitational anomaly, while simultaneously satisfying the holographic criterion in Eq.~\rr{PPHSPlusModularHolographicConditions}.

It would be interesting to learn if there is a general argument explaining the absence of \ac{cft} in the regime \rr{LSSTMGRegime},
either a direct \ac{cft} argument at the level of the conformal bootstrap along the lines of~\cite{Hofman:2008ar,Rattazzi:2010gj,Perlmutter:2016pkf},
or else a bulk argument bounding the size of higher-derivative terms in the action based on causality, extending
the methods of \it e.g. \rm~\cite{Myers:2010jv,Camanho:2014apa} to study theories
such as~\cite{Li:2008dq}.  Some work in this direction has appeared, for instance~\cite{Edelstein:2016nml}, but
the conclusions for UV-completion of effective 3D gravity theories are not yet clear.

It is likely significant that the known existing UV-complete holographic duals, have gravitational
anomaly going as $|C\ll L - C\ll R|\uu p$ with $0 < p < 1$.  The 
exponents $G\ll N\uu{-{1/3}}$ and $G\ll N\uu{-{1/2}}$ cannot correspond to coefficients of any local term in a three-dimensional bulk gravity theory,  
where the only relevant scales are the AdS length and the Planck scale.  This is not a contradiction in
the UV complete models~\cite{Maldacena:1997de,Oz:1999it}: These models have spherical factors in the geometry, $S\uu 2$ and $S\uu 3$, respectively,
where the size of the sphere is parametrically the same as the AdS scale.  It may be that this is a universal feature
holographic conformal theories with large gravitational anomalies, forced by causality and consistency of the \ac{cft}.  It would
be interesting to discover whether this is so.\footnote{We thank Kristan Jensen for suggesting the possibility that a bulk causality
constraint along the lines of~\cite{Myers:2010jv,Camanho:2014apa} may bound the parametric size of a gravitational Chern-Simons term in $D=3$.}

\heading{What is the holographic significance of anomalous nonfactorization?}

In each of the UV-complete holographic theories we have considered, the gravitational anomaly is larger than $O(1)$ but smaller than the central charge,  corresponding to an effect
smaller than classical level but larger than one-loop order.  The bulk interpretation is unclear.  Since the effect is smaller
than classical, scaling with a fractional power of $G\ll N\uu{-1}$ less than $1$, the \ac{rt} surface~\cite{Ryu:2006bv} should not show any inconsistencies at the classical level.
But it is unclear how to assign meaning to the entanglement measured by the \ac{rt} surface, since the tensor factorization of the \ac{cft} and corresponding
tensor factorization of the bulk gravity theory, appear not to be well-defined.
Though the formal manipulations~\cite{Lewkowycz:2013nqa} proving the \ac{rt} formula would seem to make sense in the bulk, at the classical level\, they appear not
to correspond to well-defined operations at the quantum level in a gravitationally anomalous \ac{cft}.  
 It is unclear, then,
how one should interpret the quantum entropies corresponding to the \ac{rt} formula or especially the quantum
corrections to it~\cite{Faulkner:2013ana}, whose size is exceeded parametrically by the magnitude of the obstruction.

The
breakdown of the assumption of tensor factorization in the boundary and in the bulk, must have real physical consequences
for the study of quantum matter coupled to gravity.  Much of the recent progress in proving model-independent properties
of quantum gravity, depends on the notion of localizability of quantum information in the boundary and in the bulk on the AdS scale,
which surely fails in its strongest form in theories with a gravitational anomaly.  However, recent developments such
as the general proof~\cite{Balakrishnan:2017bjg} of the \acl{qnec} use only the relative entropy~\cite{Casini:2008cr}, which is a coarser measure
of entanglement then a full reduced density matrix associated to each region.  Quantities such as the relative entropy,
in particular, are constructed only from the logarithm of the (relative) modular operator $\D\ll{\Psi|\Psi\pr}$, and in
therefore encodes only infinitesimally short-time behavior of the modular flow.  By 
contrast, the anomalous obstruction to the existence of a modular Hamiltonian is most clearly visible in the behavior of high powers of the modular operator,
in which a well-defined entanglement spectrum should be distinctly visible when it exists.  It would be interesting to understand how the anomalous nonfactorization manifests itself in properties of the relative entropy, mutual information, or other renormalized entanglement quantities in the gravitational dual, for instance in AdS/CFT dual pairs such as those we have discussed above~\cite{Maldacena:1997de,Oz:1999it}.

In particular, it would be interesting to consider the issue of anomalous nonfactorization in the context of "relative modular flow" ~\cite{Lashkari:2018oke,Lashkari:2018tjh}, which is intended to bear the same
relationship with to relative entropy that the ordinary modular flow has with ordinary entanglement entropy.  The primary virtue of 
relative entropy in continuum field theory and quantum gravity~\cite{Casini:2008cr} has been to give a mathematically precise renormalized version
of entanglement entropy, subtracting the nonuniversal, UV-divergent counterterms at the entangling surface while isolating universal,
UV-insensitive aspects of local quantum information of a state.  We have emphasized throughout that the anomalous obstruction to tensor factorization
is a universal, long-distance, UV-insensitive property of a \ac{cft}; it would be clarifying to see this property in the properties
of a renormalized object such as the relative modular flow discussed in~\cite{Lashkari:2018oke,Lashkari:2018tjh}, in which
UV-divergences are absent from the start.

\section{Related work}
\label{sec:related}

\subsection{Relation between our boundary construction and that of \acl{ot}}\label{UsComparedToOT}

\bigskip

%

We will here concentrate on the case where the subsystem is an interval, but the following arguments are easily generalisable to other cases.
In particular, when we write subsystem $A$ or $B$, understand them respectively as $I$, an interval, and $\bar{I}$, its complement, below.

\heading{Relations between ours and \cite{Ohmori:2014eia}}

In this paper we have emphasized the conceptual similarity between our construction and that of \ac{ot}~\cite{Ohmori:2014eia}.  
In common
the two constructions take as the starting ingredient, a definition of the tensor factorization of the state space,
\emph{i.e.}  a map from the Hilbert space on ${\cal H}\ll{A\cup B}$ to the tensor product %
Hilbert space ${\cal H}\ll A \tot {\cal H}\ll B$.
In each case, the output is a unitary, energy-preserving boundary condition. 
In both cases,
the boundary condition is obtained by a scaling limit, obtained by choosing a short distance scale $\e\lrm{short}$,
and taking the $n\uu{\rm{\underline{th}}}$ power
of a reduced density matrix $\r\ll \MacroForIOrA$, to obtain a partition function on the interval $\MacroForIOrA$ in
the limit $n\to \infty, \e\lrm{short}\to 0$ while the effective inverse temperature $(\b / L)\lrm{eff} \equiv (n\e\lrm{short} / L\lrm{original})$ is held fixed.\footnote{We will take the Renyi index to be $n$ throughout.
In what follows we will call the length of the interval to be $L$,
but this is different from what they call so in~\cite{Ohmori:2014eia}.
We will call the latter $L_{\rm OT}$, and likewise all the notations in~\cite{Ohmori:2014eia} will be called with an index ``OT''.
}

There are several differences between their construction and our own: 
\bi
\item{In the construction of~\cite{Ohmori:2014eia}, the short scale $\e\lrm{short}$ is defined, in two different constructions, as the regulator scale
of the bulk \ac{cft} itself, or the time resolution $\efac$ of the factorization map in the continuum \ac{cft},\footnote{As we will properly define in a short while, this will be defined to be the circumference of the time resolution circle, not the radius.} made finite so that the factorization
maps normalizable states of ${\cal H}$ to normalizable states of the tensor product ${\cal H}\ll A \tot {\cal H}\ll B$.
In our own construction, the \ac{cft} is defined entirely in the continuum and there is no bulk regulator; and the resolution time $\efac$ of the factorization
decouples entirely from the path integral.  The nonuniversality of the 
factorization data arising from these two scales, is transferred to the scale $\e$ of the inverse temperature of the thermal ensemble of the full system.}

\item{Our assumption on the nature of the factorization map is weaker than that made
in~\cite{Ohmori:2014eia}.  In the present work, we do \rwa{not} necessarily assume neither that the \ac{cft} has a lattice realization, nor
that the factorization map is defined by a path integral with 
a boundary condition of any kind.  Rather, we assume only that the factorization map respects locality of the operator algebra (in the
sense defined precisely in Appendix~\ref{LocalityCriterionPreciseDefinition}) inserted at the entangling
points at the boundary of subregion $\MacroForIOrA$.}
\item{Our construction is done entirely in a single conformal frame.  This is useful because our aim is to demonstrate the nonexistence of
well-defined tensor factorizations in gravitationally anomalous theories.  Deriving a boundary condition in the coordinate frame to define
the \ac{cft} in the first event, bypasses the subtleties associated with coordinate transformations in gravitationally anomalous \ac{qft}.}
\ei

Despite the differences, we shall show that
our construction is not only analogous to that of~\cite{Ohmori:2014eia}, it produces literally the same boundary condition in the scaling limit $n\to\infty$, for
any given UV-complete factorization data.  

\heading{The short-distance scale(s) in our construction and in \ac{ot}}

Let us review the two roles of a short-distance
scale $\e\lrm{short}$ used in the two distinct ways of defining the factorization map ${\cal M}:{\cal H}\mapsto {\cal H}\ll A \tot {\cal H}\ll B$
in~\cite{Ohmori:2014eia}. (There, both scales are referred to as "$\e$",
while in the present paper we have reserved $"\e"$ for the inverse
temperature of the thermal ensemble on $A\cup B$.) In the first part of the paper, the authors of~\cite{Ohmori:2014eia}, emphasize that the factorization map can be understood entirely in the continuum, with the scale (which they call $"\e"$ and we will call $\efac$) being the resolution time-scale
of the factorization, introduced to make the normalization of the map finite (See the Introduction and the Appendix \ref{PreciseDefinitionOfLocalityForQuantumInfoStuff} for more details.). 
Although ``$\epsilon$'' in \cite{Ohmori:2014eia} was defined slightly differently from $\epsilon_f$ by a numerical factor, we decide here that the precise definition of our $\epsilon_f$ is the circumference (not the radius) of the time-resolution circle, in the continuum theory.

Later in~\cite{Ohmori:2014eia}, the authors consider examples of lattice theories, where the UV scale is simply the lattice spacing, (which they also call $"\e"$ and
which we shall call $\e\ll{\textsc{uv}} = \L\uu{-1}\ll{\textsc{uv}}$), with
the factorization defined directly \it via \rm some boundary conditions.  

Taking sufficiently large powers $n$ of the density matrix makes the circumference of the circle surrounding the entangling surface,
grow as $n\e\ll{\textsc{uv}}$.  When $n\e\ll{\textsc{uv}} \gtrsim \LCFT\uu{-1}$, the entangling surface can be approximated by
a continuous boundary.  Then increase $n$ to some sufficiently large value $n\ll 0$ such that 
the length $n\ll 0 \e\ll{\textsc{uv}}$ is much larger than $\LCFT\uu{-1}$ but still small compared to the size $L$ of the subregion:
\bbb
n\ll 0 \cc
\e\ll{\textsc{uv}} = n\ll 0\cc \L\uu{-1}\ll{\textsc{uv}} \equiv\efac \muchgreaterthan \LCFT\uu{-1},
\een{EquationWithNSubscriptZeroWhichIWouldLikeToGetRidOf} 
 where $\LCFT^{-1}$ (as defined below \rr{EquationAboveWhereWeMentionTheCFTScaleLambdaCFT}) is the distance
scale at which the lattice theory is approximated by its conformal continuum limit to any desired approximation; one can
do this while still keeping the distance scale $\LCFT^{-1}$ much smaller than the size $L$ of the interval $\MacroForIOrA$.

At this scale, the lattice structure is washed out, and the factorization map can be understood as a path integral with a boundary
of size $\efac$. One can then further increase the power of $\r\ll I$ in the trace,
increasing $n$  by a factor of $n\lrm{OTC}$ to $n\lrm{OTL}\equiv n\ll 0 \cc n\lrm{OTC}$ to obtain a boundary of any desired macroscopic size $ \b\lrm{OT} \equiv  n\lrm{OTC}\efac =  n\lrm{OTL} \e\ll{\textsc{uv}} $ (No factor of $2\pi$, since $\epsilon_f$ is the circumference, not the radius.).

The number $n\lrm{OTC}$ is the number of times the path integral must be replicated (the power to which the density matrix must be raised) from the starting point of
a Hilbert space factorization defined in the continuum, in the construction described in the first part of~\cite{Ohmori:2014eia}; the number  $n\lrm{OTL} = n\ll 0 n\lrm{OTC}$
is the number of times the path integral must be replicated (the power to which the density matrix must be raised) from the starting point of
the factorization procedure defined on the lattice, in the concrete constructions described in the latter part of~\cite{Ohmori:2014eia}

%

%
{As we saw above, we can forget about the lattice theory and work entirely in the continuum once we take the high enough power $n_0$ of the reduced density matrix of the original lattice theory.
The continuous geometry on which the theory, now a \ac{cft}, lives, can be conformally transformed ($z\mapsto -\log z + \log(L-z)$ in units where $\frac{\epsilon_f}{2\pi}=1$) to a simple thermal cylinder, so that the $n_0$-th reduced density matrix (of the lattice theory) can be computed as the thermal partition function of the \ac{cft} on an interval with length and temperature being
\begin{align}
\ell_{\rm OT}=\frac{\epsilon_f}{\pi}\log\left(\frac{2\pi L}{\epsilon_f}\right) \quad\text{and}\quad\beta_{\rm OT}=\epsilon_f 
\end{align}
Likewise, when we increase the Renyi index to $n_{\rm OTL}$ in the lattice theory, this amounts to taking the $n_{\rm OTC}$-fold branched cover in the continuum theory above, and so after the same conformal transformation, the geometry is again a thermal cylinder.
In other words, the $n_{\rm OTC}$-th reduced density matrix of the lattice theory is the same as the 
$n_{\rm OTL}$-th reduced density matrix of the theory in the continuum, and so after the same conformal transformation it
 can be computed as the thermal partition function of the \ac{cft} on an interval with length and temperature 
\begin{align}
\ell_{\rm OT}=\frac{\epsilon_f}{\pi}\log\left(\frac{2\pi L}{\epsilon_f}\right) \quad\text{and}\quad\beta_{\rm OT}=n_{\rm OTL}\epsilon_f 
\end{align}
}
The authors show, by explicit calculation, that the boundary condition obtained by this coarse-graining
is unitary -- \it i.e. \rm satisfying the Cardy condition -- and nonuniversal, with the boundary state $\kket{{\tt bdy}}$ and spectrum of the Hamiltonian on the interval
depending on the choice of UV-completed factorization data.

We now list \rwa{all} the scales
playing a role in this section and related discussions elsewhere,
and note the logically necessary hierarchies in magnitude between some
of them:
\bbb
\L\ll{\textsc{uv}}\uu{-1} =  \e\ll{\textsc{uv}} \lesssim \{\efac,\L\uu{-1}\lrm{CFT}  \}\muchlessthan \e  \muchlessthan\{ L \cc , \cc \b = n \e \}
\een{BigScaleMultipleHierarchyDisplay}
In our own construction the scales $\L\lrm{UV}\uu{-1}, \L\lrm{CFT}\uu{-1}, \efac$
are irrelevant: we have taken them all to zero, with the nonuniversality
originating in the boundary state, transferred to the thermal scale $\e$. 
The scale $\efac$ cannot meaningfully be made smaller than the lattice scale
$\e\ll{\textsc{uv}}$, but it can be either much longer than 
the scale $\L\lrm{CFT}$ (as in the discussion in the earlier part of~\cite{Ohmori:2014eia}) or much shorter than $\L\lrm{CFT}$ (as in the detailed calculations
in the later part of~\cite{Ohmori:2014eia}).

\heading{Conformal transformation between our construction and that of~\cite{Ohmori:2014eia}}

\begin{figure}[t]
 \begin{center}
 \begin{overpic}[width=0.8\columnwidth]{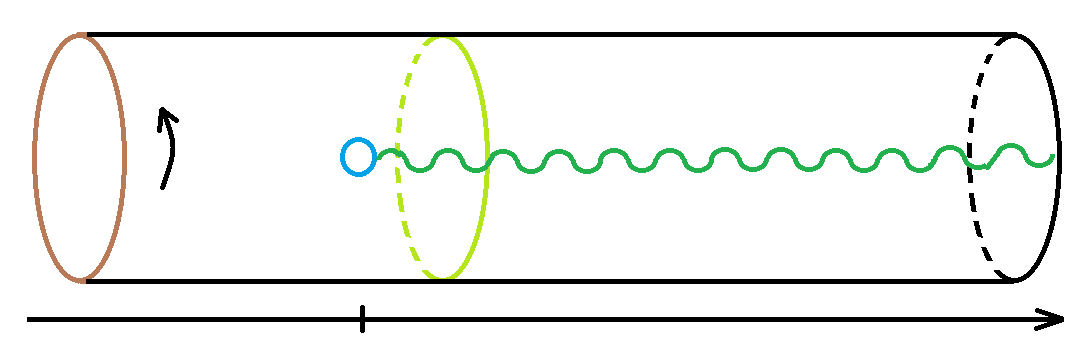}
 \put(17,17){$\sigma\times \frac{\epsilon}{2\pi}$}
 \put(100,0){$\tau\times \frac{\epsilon}{2\pi}$}
 \put(28,-3){$-L/2$}
\end{overpic}
 \end{center}
 \caption{This is (half of the) quotient geometry for our construction, where we cut the whole geometry of figure \ref{fig:quotient-double} in half on the right edge. The brown circle is of circumference $\epsilon$ as we regularize the trace using heat-kernels. Also, the blue circle is of  circumference $\efac$ in~\cite{Ohmori:2014eia}, but this scale can be shown  via conformal transformation to decouple in our finite-temperature construction.
The circle in yellow-green is the Euclidean-time circle whose circumference defines the inverse temperature of the thermal ensemble on
the full spatial slice $\Sigma = A \cup B$, on which the boundary states live in our construction. After an open-closed modular transformation, this can be used in the dual description of Sec. \ref{sec:factorization-implies-boundary}~\cite{hep-th/0411189}.
}
 \label{quo}
\end{figure}

In this paper, we have put the theory on a cylinder of radius $\epsilon$, and discussed properties of traces of n$\uth$ powers of the reduced density
matrix $\r\ll I$ of the interval.  As reviewed in sec. \ref{sec:CardyCalabrese}, these traces can be realized as path integrals on
a singular Riemann surface with branch cuts terminating on the entangling points.
In the definition of these path integrals, the UV divergence at the conical singularity is resolved by the (non-universal~\cite{Ohmori:2014eia}) data of the factorization map.
In a conformal theory with $C\ll L = C\ll R$, this computation is equivalent to the partition function for the $n$-branched cover of a cylinder.
Say we want to compute the trace of the n$\uth$ power of the reduced density matrix for an interval $\MacroForIOrA$ of length $L$ when the \ac{cft} full is at inverse temperature $\e$.
In this case the geometry we are considering is the one in Figure~\ref{quo} (for reasons of legibility we show only half the geometry),
and with a conformal transformation we can set the radius to be $2\pi$ and the length of $\MacroForIOrA$ to be $2\pi L/\epsilon$.

In our construction, we use the boundary state living at the two ends of the interval to compute the entanglement entropy of an interval -- This is the same as computing the closed string amplitude like the following
\begin{align}
\left\langle{A\left|e^{-LH_{\rm closed}}\right|B}\right\rangle
\end{align}
with the inverse temperature being $n\epsilon$.
This corresponds to the Fig. \ref{quo} -- As one can see, the circumference of the circle (indicated in light green in the figure) on which the boundary states live is $n\epsilon$, being promoted $n$ times because of the presence of the branch cut.
This closed-string amplitude is equal to the open string spectrum,
\begin{align}
\Tr_{AB}\left[e^{-n\epsilon H_{AB}}\right].
\end{align}

We now do a conformal transformation into a planer geometry, which maps the partition function of a cylinder with circumference $\beta$ and length $L$ to
\begin{align}
\Tr_{AB}\left[\left(e^{-\frac{\pi \beta}{L}}\right)^{\Delta}\right]=\sum_{\Delta_i}e^{-\frac{\pi \beta \Delta_i}{L}}
\end{align}
In the case of our interest, therefore, we have
\begin{align}
\Tr_{AB}\left[\left(e^{-\frac{\pi n\epsilon}{L}}\right)^{\Delta}\right]=\sum_{\Delta_i}e^{-\frac{\pi n\epsilon \Delta_i}{L}}
\end{align}
where $\Delta$ is the dilatation operator and $\Delta_i$ are its eigenvalues.
Note that this gives us the spectrum of the modular Hamiltonian, and so this will be something we compare between the construction of ours and that of \cite{Ohmori:2014eia}.
Proving that we get the same spectrum between the two proves that they are equivalent.

Turning to the construction of \cite{Ohmori:2014eia}, they use the boundary condition of the resolution circle to compute the same entanglement spectrum.
Let us point out again that we compute the same thing in two different ways, which look completely different with one another. 
This is why we would like to check if these two ways of computing the entanglement spectrum are really consistent.

In the construction of \cite{Ohmori:2014eia}, we first introduce a resolution circle at the boundary, which resolves the tip of the conifold introduced by the branch cut.
We set the size of the two resolution circles to be both $\epsilon_f$, but this will again be magnified $n$ times to become $n\epsilon_f$ because of the branch cut (See Fig. \ref{quo}). 
The centres lie at $\tau_{A}=-\frac{\pi L}{\epsilon}$ and $\tau_{B}=\frac{\pi L}{\epsilon}$ and both at $\sigma=0$, so points on the circles should be parametrized using parameter $\theta_A$ and $\theta_B$ as 
\begin{align}
z_A(\theta_A)=-\frac{\pi L}{\epsilon}+\frac{\epsilon_f}{\epsilon}e^{i\theta_A}
\end{align}
and
\begin{align}
z_B(\theta_B)=\frac{\pi L}{\epsilon}-\frac{\epsilon_f}{\epsilon}e^{i\theta_B}
\end{align}
where $0<\theta_{A,B}<2\pi n$, again because of the branch cut.

In order to make connection to \cite{Ohmori:2014eia}, what we now do is a conformal transformation that maps the cylinder in Fig. \ref{quo} into a plane,
\begin{align}
z=\tau+i\sigma\mapsto w=e^{z} 
\end{align}
where $\tau$ and $\sigma$ are dimensionless as they are defined to be $2\pi/\epsilon$ times the actual length.
After this conformal transformation, the centre of the two resolution circles map to 
$w=e^{\pm \frac{L\pi}{\epsilon}}$, with the radius of the hole being $\frac{\epsilon_f}{\epsilon}e^{\pm \frac{L\pi}{\epsilon}}$ at leading order in $\epsilon_f/\epsilon \muchlessthan 1$.

Now, one can do the conformal transformation
\begin{align}
w\mapsto \xi=-\log\left(w-e^{- \frac{L\pi}{\epsilon}}\right)+\log\left(e^{\frac{L\pi}{\epsilon}}-w\right)
\end{align}
which sends $z_A$ and $z_B$ to
\begin{align}
\xi_A(\theta_A)=\frac{2L\pi}{\epsilon}-\log\left(\frac{\epsilon_f}{\epsilon}\right)+i\theta_A
\end{align}
and 
\begin{align}
\xi_B(\theta_B)=\log\left(\frac{\epsilon_f}{\epsilon}\right)+i\theta_B
\end{align}
As we can see, this conformal transformation maps the resolution circles to the left and the right boundary of a new cylinder in Fig. \ref{quo2}.

\begin{figure}[t]
 \begin{center}
 \begin{overpic}[width=0.8\columnwidth]{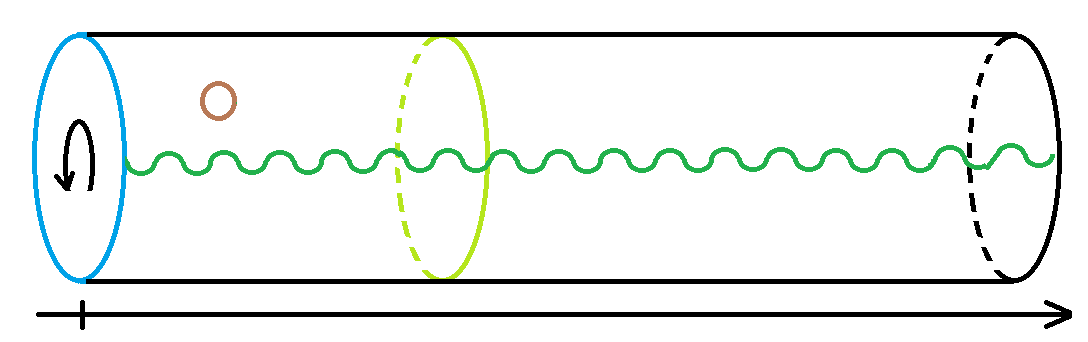}
 \put(5,11){$2\pi$}
 \put(100,1){$\Re \zeta$}
\put(1,-2){$-\ell_{\rm OT}/2$}
\end{overpic}
 \end{center}
 \caption{This is half of the quotient geometry in the construction of~\cite{Ohmori:2014eia}, where we again cut the whole geometry in half on the right edge. After conformal transformation from Fig. \ref{quo}, the radius of the cylinder is $2\pi$ and the length $2\pi L/\epsilon$.
}
 \label{quo2}
\end{figure}

The branch cut runs from the left to the right boundary as shown in Fig. \ref{quo2}, and so the circumference of the cylinder becomes $2\pi n$. 
The length of the cylinder is simply
\begin{align}
\Re(\xi_B-\xi_A)=\frac{2L\pi}{\epsilon},
\end{align}
in the limit where $\epsilon_f\muchlessthan \epsilon \muchlessthan L$, as well as in the limit $\log\left(\frac{\epsilon_f}{\epsilon}\right)$ is negligible compared with $\frac{2L\pi}{\epsilon}$.
Incidentally, in order to recover the dimensionful quantity, one should multiply everything by $\frac{\epsilon}{2\pi}$, which we set to be $1$ when we defined $\tau$ and $\sigma$.
Note that this expression is different from $\ell_{OT}$ introduced above, since they dealt with flat space, while we started from the cylinder by introducing some temperature to regulate the boundary states.

Now, it is easy to compute the entanglement spectrum from this argument, because we see that this is simply the open-string spectrum of the cylinder with circumference $2\pi n$ and length $\frac{2L\pi}{\epsilon}$.
After recovering the dimensionful factor, the circumference is $n\epsilon$ and the length $L$ --
This is exactly the same geometry as that of our construction.
The entanglement spectrum we compute in this way is
\begin{align}
\Tr_{AB}\left[\left(e^{-\frac{\pi \times 2\pi n}{2L\pi/\epsilon}}\right)^{\Delta}\right]=\sum_{\Delta_i}e^{-\frac{\pi n\epsilon \Delta_i}{L}}
\end{align}
which gets us the same result (or more specifically, the same Boltzmann factor) as the previous construction of ours, proving the consistency of the two ways of computing the Renyi entropy using boundary state formalism.

\begin{figure}[t]
 \begin{center}
 %
 %
\begin{overpic}[width=0.4\columnwidth]{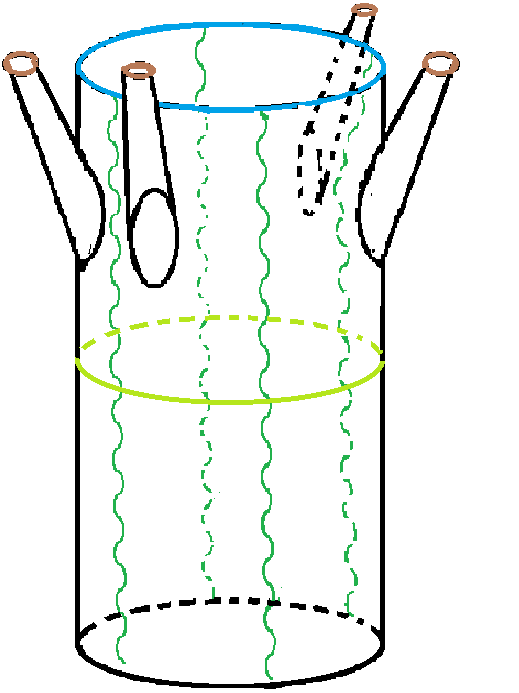}
\end{overpic}
 \end{center}
 \caption{This is the octopus geometry, half of the geometry of the $n$-fold cover of the infinite cylinder with two $n$-order twist operators
 inserted; in the figure we take $n$ to be $4$. Green wavy lines indicate the positions of what used to be the branch cuts in the original geometry.
 The replica transformation acts as a $2\pi/n$ rotation of the large cylinder, where again $n=4$ in this figure.}
 \label{octo}
\end{figure}

We would like to reiterate that the equivalence of these two methods are nontrivial, as they are not simply related by a simple rescaling map, in fact, the conformal transformation proving the equivalence was very complicated.
Nontrivial as it is, though, one can better see this equivalence at the qualitative level by thinking about the covering space, not the quotient, of the geometry we are interested in.
The geometry is shown in Fig. \ref{octo}, looking like a cylinder with many legs.
The ends of this cylinder are what used to be the resolution circles, on which the boundary states in the \cite{Ohmori:2014eia} construction lives. 
Past many legs (of which there are as many as the Renyi index $n$), which used to be the left and the right infinity of Fig. \ref{quo}, we have what used to be the boundary states in our construction.
Now, by capping off the legs with vacuum, it is immediate that our boundary states and the boundary states of \cite{Ohmori:2014eia} are equivalent.
Together with the explicit computation of the length of the cylinder, the equivalence is now apparent, as promised.

To summarize, a nontrivial conformal transformation maps between our and \cite{Ohmori:2014eia}'s construction of the entanglement spectrum, finally checking the consistency of our argument.

\def\OldSectionDefout{ %

\begin{itemize}
\item In our construction the computation is tantamount to using the
  transfer matrix that runs from left to right on subsystem $\MacroForIOrA$; the
  result then becomes schematically
  \begin{equation}\label{MWSumOverIntervalOpDimensions}
    \redd{ \sum_{\Delta_{\MacroForIOrA}}\exp\left[-\frac{2\pi L\Delta_{\MacroForIOrA}}{n\epsilon}\right].} 
   \llsk\blue{\leftarrow} \purple{\rm \huge{NO!}}
  \end{equation}
  \shg{I think this should be:
  \begin{equation}\label{MWSumOverIntervalOpDimensions}
 \sum_{\Delta_{\MacroForIOrA}}\exp\left[-\frac{\b\lrm{eff}\Delta_{\MacroForIOrA}}{L/\pi}\right] =  \sum_{\Delta_{\MacroForIOrA}}\exp\left[-\frac{\pi\b\lrm{eff}\Delta_{\MacroForIOrA}}{L}
 \right]  =  \sum_{\Delta_{\MacroForIOrA}}\exp\left[-\frac{n\pi \Delta_{\MacroForIOrA}}{\e L}
 \right].
  \end{equation}
  }
\item The view of~\cite{Ohmori:2014eia}, on the other hand, is the following.
Regulate the branch point using a hole of radius $\efac$,
and make a conformal transformation such that the hole itself becomes a state as in Figure~\ref{quo2}. The conformal transformation that does it is a combination of $z=\tau+i\sigma\mapsto w=e^{z}$ and $w\mapsto \zeta=-\log (w-\a^{-1})+\log(\a-w)$, where $\a\equiv \exp\left[\pi \cc L/\epsilon\right]$.
\end{itemize}

\begin{figure}[t]
 \begin{center}
 \begin{overpic}[width=0.8\columnwidth]{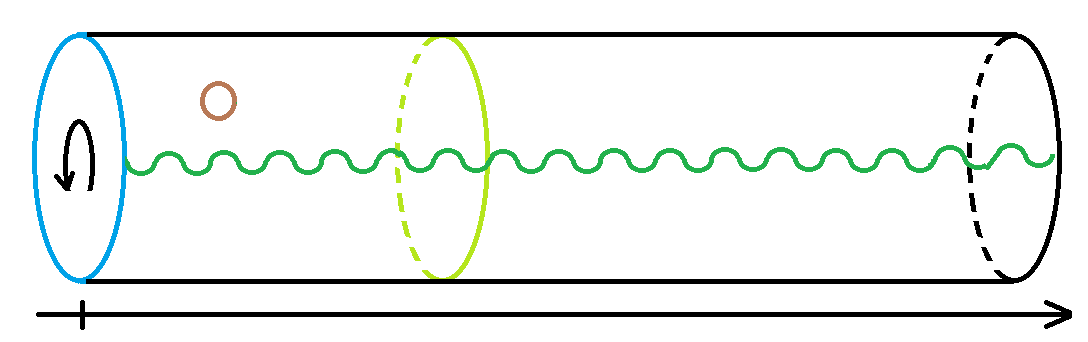}
 \put(5,11){$2\pi$}
 \put(100,1){$\Re \zeta$}
\put(1,-2){$-\ell_{\rm OT}/2$}
\end{overpic}
 \end{center}
 \caption{This is half of the quotient geometry in the construction of~\cite{Ohmori:2014eia}, where we again cut the whole geometry in half on the right edge. After conformal transformation from Fig. \ref{quo}, the radius of the cylinder is $2\pi$ and the length $2\pi L/\epsilon$.
}
 \label{quo2}
\end{figure}

An explicit computation shows that the length of the cylinder in the Ohomori-Tachikawa frame is just
\begin{equation}
 \ell_{\rm OT}=\displaystyle\frac{2\pi \cc L}{\epsilon}  \pqty{1+\order{\frac{\epsilon}{L}\log(\frac{\efac}{\epsilon})}}, 
\label{FormulaWhichLooksNothingLikeTheCorrectFormla}\end{equation}
\shg{Are we certain this formula, eq. \rr{FormulaWhichLooksNothingLikeTheCorrectFormla}, is consistent with the formula OT give for $\ell\lrm{OT}$ in their paper below their eq. \blue{(2.2)}, which is
\bbb
\ell\lrm{OT} = 2\cc \e\lrm f\cc {\rm log}[L / \e\lrm f]
\eee
which I have used in the paragraphs above?}
which is the same as the length of $\MacroForIOrA$ that we had found above.

} %

\subsection{Relation to the phenomenon observed by \acl{achp}}
 
 As this paper was being completed, another paper appeared studying a free \ac{cft} with $c\ll L = 0$ and $c\ll R = 1$ by \ac{achp}~\cite{Arias:2018tmw}, 
 noting the
 impossibility of constructing reduced density matrices in a local way in
 that theory.  Their calculation illustrates concretely by example the general proof
 given in the present paper and in~\cite{WatanabeThesis:2015,SHSeminarsIncludingIPMU2015}.  In the Conclusions,
 the authors pose
 the question, ``It would be interesting to understand why the fermion modular Hamiltonian is quasilocal while the one
of the current is completely nonlocal.'', which the present paper settles conclusively.
 
 Their work differs from our own in one important respect; the authors 
 refer in many instances to ``the'' modular Hamiltonian of a subregion
 of the \ac{qft}.  As originally pointed out in~\cite{Ohmori:2014eia} and 
 as we have emphasized throughout, there is simply no such thing as ``the''
 modular Hamiltonian of a subsystem in a continuum theory.  A local definition
 of the modular Hamiltonian is non-unique when it exists~\cite{Ohmori:2014eia} and
 does not always exist.  In two dimensions, there always exist nonlocal
 definitions of the tensor factorization and reduced density matrix, exemplified
 by the Holzhey--Larsen--Wilczek construction~\cite{Holzhey:1994we}, generate well-defined factorization
 data, but violate the locality criterion as define in Appendix \ref{LocalityCriterionPreciseDefinition}.  As a result,
 the operator ${\cal F}[{\cal M}\lrm{HW}]$ is not a product of almost-local operators
 at the boundary of the subregion $A$, but a superposition of almost-local operators,
 so that the image $\ket{\Psi\pr} \equiv {\cal F}[{\cal M}] \cc \ket{\Psi}$ of a state $\ket{\Psi} \in {\cal H}\ll\Sigma$
 under ${\cal F}[{\cal M]}$, has nonlocal correlations between quantum fields
 at macroscopically separated boundaries of subregion $A$, that were not present in
 the original state $\ket{\Psi}$.  The result is that the modular flow generated by
 $\r\ll A$ always involves macroscopic nonlocality, particularly teleportation of
 information from among boundary components. In a gravitationally anomalous theory,
 macroscopically nonlocal definitions of the tensor factorization, such as those given in~\cite{Holzhey:1994we}, are
 the only ones available; this is the underlying reason for the otherwise-puzzling difference between the
 behavior of the modular flow in the cases of chiral \it vs. \rm nonchiral free theories, as
 noticed recently in~\cite{Arias:2018tmw}.  
 
\subsection{Cobordism classes, domain walls, and universe selection rules in quantum gravity }

Recent work on the ``swampland'' program~\cite{Vafa:2005ui} has investigated the conditions in which a consistent theory of quantum gravity can allow superselection sectors.
That is, one asks whether two vacua of a quantum gravity theory may be separated by a finite-tension domain wall, or whether they
are unconnectable by such a configuration.  In particular, the papers~\cite{McNamara:2019rup,McNamara:2020uza}
offered evidence that any two vacua of a consistent quantum gravity theory must be dynamically connectable by a domain
wall, generalizing the older notions of the stringy duality web of M-theory.\footnote{For earlier work on the question of dynamical connectedness of the solution space of string theory beyond the supersymmetric duality web, see \emph{e.g.}~\cite{Srednicki:1998mq,Sen:1998tt,Bergman:1999km,Hellerman:2004zm,Hellerman:2004qa,Schnabl:2005gv,Hellerman:2006ff,Hellerman:2006hf,Hellerman:2007fc,Hellerman:2007zz,Hellerman:2010dv}, as well as recent developments~\cite{Kaidi:2020jla} along related lines.}  In the context of AdS quantum gravity, this
issue was explored though interfaces between \ac{cft} and their correspondence with domain walls in the holographic dual.  
Through this correspondence it was observed (using the proof originally given in~\cite{WatanabeThesis:2015,SHSeminarsIncludingIPMU2015}
  on the absence of boundaries for theories with
 nonzero gravitational anomaly, generalized to the case of interfaces)
 that a difference in the gravitational anomaly coefficient of two boundary CFT${}\ll 2$ is an obstruction to connecting 
 the two corresponding vacua of the holographic gravity theory in AdS${}\ll 3$.

\subsection{Misc.}

Many of the results in the present paper have already appeared in the Master's thesis of one of the authors~\cite{WatanabeThesis:2015},
particularly the no-go theorem for boundary conditions in gravitationally anomalous theories, which was
later transcribed into the language of anomaly polynomials in ref.~\cite{Jensen:2017eof}.  Additionally, a paper appeared earlier~\cite{Han:2017hdv} \footnote{We thank Yuji Tachikawa for bringing this reference to our attention.} 
 which independently drew conclusions overlapping partially with our own.  Also, recently a paper appeared~\cite{2012.15861} deriving further anomalous obstructions
 to the existence of boundary conditions for various 't Hooft anomalies in diverse dimensions, drawing on the theorems that appeared originally
 in~\cite{WatanabeThesis:2015,SHSeminarsIncludingIPMU2015}.

\section{Conclusions}

In this paper we have given a general and elementary proof that the Hilbert space of a two-dimensional quantum field
theory can never be factorized into a tensor product of Hilbert spaces supported in disjoint spatial regions,
in a way that respects locality of the operator algebra on macroscopic scales, if the gravitational anomaly $C\ll L - C\ll R$ is nonvanishing.
This gives a broad generalization of the (two-dimensional) Nielsen-Ninomiya theorem~\cite{Nielsen:1980rz,Nielsen:1981xu,Nielsen:1981hk} that is applicable to general unitary interacting \ac{qft}s in two dimensions, while the very original proof was only applicable to free fermionic systems.

Our construction applies to nonconformal as well as conformal quantum field theories (see the discussion in sec. \ref{BoundaryAndBulkRGFlow}
of the Appendix), so long as they are unitary and
their long-distance dynamics is relativistic.  This nonfactorization is robust and produces meaningful 
physical consequences for the theory at long distances, far from the boundary between regions, the so-called
entangling surface.
This provides a generalization of the work of and Nielsen and Ninomiya on chiral fermions~\cite{Nielsen:1980rz,Nielsen:1981xu,Nielsen:1981hk} for we have shown that it is in general impossible to obtain a gravitationally anomalous theory as the continuum limit of a local lattice theory with finite entropy per site.
Most striking is the impossibility of uniformly suppressing correlations in a pure state, between any two regions, 
regardless of the amount of available energy.

We have shown that any local definition of the partial trace, produces a Hamiltonian on a region with boundary, by taking a scaling
limit of the partial trace of the modular Hamiltonian for the full system at high temperature.  The basis of the construction
is that while a modular Hamiltonian is nonlocal in general, a high-temperature modular Hamiltonian must be local on scales long
compared to the inverse temperature $\e$, if the partial trace is defined in such a way as to respect locality of the operator algebra.

We have noted that much interesting work on the quantum information theory of gravity via bulk-boundary holography, has
been based on the assumption of factorizability of the Hilbert space, an assumption which is violated parametrically
in the central charge, even in some known examples of holographic dual pairs~\cite{Gauntlett:2000ng,Oz:1999it}. It would be interesting
to understand the consequences of anomalous nonfactorization of the \ac{cft} when interpreted in the holographic dual.

The derivation of an energy-conserving, unitary boundary condition from a consistent set of tensor factorization data, plays an
essential role in the no-go theorem for the localization of quantum information.  This connection has been noted previously~\cite{Ohmori:2014eia}; we have emphasized the analogy with that work, and clarified the relation of our own construction to the one appearing in~\cite{Ohmori:2014eia}.
There, the authors emphasized the non-uniqueness of the UV-completing terms 
as the origin of the non-uniqueness of the corresponding boundary state.  In the current context, we conclude that in addition
to non-uniqueness, there is a corresponding issue of nonexistence in some cases; specifically if the gravitational anomaly
is nonvanishing.  Our own derivation also simplifies theirs in certain respects, in particular by constructing the boundary condition from
factorization data, entirely in a single conformal frame.

\heading{Future directions}

We would also like to note some possible interesting future directions to follow up on the results presented here.

First, it would be valuable to understand the consequence towards quantum information theoretic proof towards $c$ or $a$-theorem or \ac{qnec} where it seems like the fundamental assumption is to use the factorizability of the Hilbert space.
If it turns out that the assumption is absolutely important in proving those results,
it might happen that one can find a ``counterexample'' of \ac{qnec} in
gravitationally anomalous theories, as it still lacks a proof using minimal field-theoretic "bootstrap"-like considerations, in contrast
to the proof of the \ac{anec}~\cite{Borde:1987qr,Roman:1986tp,Roman:1988vv,Hartman:2016lgu}, which does use only such minimal ingredients.

It would be good to give a simpler derivation of the no-go principles derived here that would bypass the technical issues of counterterms and renormalization, by considering the "relative modular flow" idea of
\cite{Lashkari:2018oke,Lashkari:2018tjh}.  It would also be good to understand whether there are any implications for the properties of simple renormalized
measures of entanglement such as relative entropy or mutual information.

Quantum information theory has also been important in understanding quantum gravity and the black hole information paradox. 
Recently, the simplest version of the paradox, the so-called ``firewall'' paradox, concerning the von-Neumann entanglement entropy, has been resolved purely at the level of effective general relativity, by consistenly treating exotic saddle-points of the multi-replica theory \cite{Almheiri:2019qdq,Penington:2019npb,Almheiri:2019psf,Almheiri:2020cfm}.
More broadly, however, one expects that the complete resolution of the paradox entails the consistency conditions of UV-complete quantum gravity.
It might be the case that the $1+\epsilon$ power of the reduced density matrix, \it i.e., \rm the entanglement entropy, encodes properties of low-energy GR, while large $\epsilon$ encodes the high-energy physics.
One may hope the present paper can be used as a model of how large-order Renyi entropies may contain information about
UV consistency conditions of quantum gravity more generally.

\def\ExtendedAnomalyDefout{
\shg{After the recent paper from last week, is this paragraph still true?  Didn't they do some things like this? If so, we should remove this paragraph.}

\redd{Aside from interpreting our result holographically. 
We would also like to point out the obvious fact that there will be no obstructions from pure gravitational anomalies in odd spacetime dimensions, where there are no such things.
However, it could be interesting to note that there can be discrete $p$-form anomalies in those dimensional theories, which might or might not serve as an obstruction to tensor factorization.
In particular, it should be very interesting to find, if any, a three-dimensional
continuum theory which can be proven to lack localizable quantum information, boundary conditions with various properties, or local lattice regulators.}
}

\section*{Acknowledgments}
{
  \sffamily \begin{small}
The authors acknowledge several people for providing essential input, including Daniel Harlow, Kristan Jensen, Alexei Kitaev, Hirosi Ooguri, Juan Maldacena, Xiao-Gang Wen.  We also acknowledge crucial discussions with Leonard Susskind and
Ning Bao on the behavior of quantum information in the limit of large-order replication, which motivated the factorization-to-boundary
construction proposed in this note.  We are particularly grateful to Xi Yin and Ibrahim Bah for pointing out counterexamples to a proposed holographic interpretation of our \ac{cft} results
in an earlier version of the draft, and to the Harvard Center for the Fundamental Laws of Nature and the Department of Physics at UCSD, respectively, for hospitality while the significance of those counterexamples
was being understood.  The work of SH~is supported by the
World Premier International Research Center Initiative (WPI Initiative), MEXT, Japan; by the JSPS Program for Advancing Strategic International Networks to Accelerate the Circulation of Talented Researchers; and also supported in part by JSPS KAKENHI Grant Numbers JP22740153, JP26400242. 
DO acknowledges support  by the NCCR 51NF40-
141869 “The Mathematics of Physics” (SwissMAP). 
The work of MW is supported by JSPS Research Fellowship for Young Scientists.
The work of MW is also supported by
the Foreign Postdoctoral Fellowship Program of the Israel Academy of Sciences and Humanities and by an Israel Science Foundation center for excellence grant (grant number 1989/14) and by the Minerva foundation with funding from the Federal German Ministry for Education and Research.
SH~also thanks the Burke Institute at Caltech, the Harvard Center for the Fundamental Laws of Nature, the Stanford Institute for Theoretical Physics, the Aspen Center for Physics, and the Department of Physics at UC San Diego for hospitality while this work was in progress.
\end{small}
}

\appendix

\section{Local factorization maps and renormalized tensor products}\label{PreciseDefinitionOfLocalityForQuantumInfoStuff}

In our no-go theorem, we have assumed a rather strong property for the notion of a tensor factorization of the Hilbert
space, namely that the factorization respect the locality of the operator algebra.  This criterion has an intuitively obvious 
meaning, and is satisfied by most but not all procedures for defining the factorization.  In this Appendix we now aim
to be more explicit and concrete regarding this criterion.

We shall also independently give a precise criterion for the "renormalized tensor product" $\tot$, which generalizes the usual notion of the tensor product
to allow non-factorized short-distance structure at the entangling surface.  We shall then note that any generalized notion factorization of a \ac{qft}
described by the general tensor product $\tot$, gives a map ${\cal M}:{\cal H}\ll{A\cup B}\mapsto {\cal H}\ll A \tot {\cal H}\ll B$ satisfying the locality
criterion as we define it in this Appendix.

\subsection{A locality criterion for tensor factorizations}\label{LocalityCriterionPreciseDefinition}

\heading{Motivation}

We want to define a locality criterion for the factorization map guaranteeing that
the splitting of the space into pieces and reconnection of those pieces into the full space over a timescale $\efac$, does not introduce any correlations that were absent previously, on scales longer than $\efac$.  
    This is the criterion that is satisfied by the boundary-condition construction of~\cite{Ohmori:2014eia}, and is
    inherited from a factorization derived from a lattice regulator.  But we wish to define the criterion
    more generally.  On the other hand we would like to make the definition narrow enough to exclude the tensor factorization
    definition of~\cite{Holzhey:1994we}, for instance, which does introduce correlations on the scale of the size of the region itself.

\heading{The general idea, and some subtleties}

The broadest definition would simply be to say
that tensor factorization is defined as an isomorphism ${\cal M}: {\cal H} \mapsto {\cal H}\ll A \tot {\cal H}\ll B$ that
respects locality, in the sense that the operator ${\cal F}[{\cal M}] \equiv {\cal M}\cc {\cal M}\dag : {\cal H}\mapsto {\cal H}$ commutes with local operators
away from the entangling surface.  This is the right idea, but we need to refine the definition to take two points into account:
 \bi
  \item{The entangling surface itself is a geometry that may have a size of $O(1)$ on the infrared scale; in the two-dimensional case,
 the entangling surface in finite volume is always made up of more than one component, so the geometry of the entangling surface is
 macroscopic, the size of the subregion.  
 We ought to restrict our definition
 so that the operator ${\cal F}[{\cal M}]$ does not introduce long-range correlations 
 between regions near different parts of the entangling surface.}
 \item{A factorization map ${\cal M}$ which
 is precisely local, \emph{i.e.} such that ${\cal F}[{\cal M}]$ is an exactly
 local operator or product of exactly local operators at the
 boundaries of $A$, never maps normalizable states
 to normalizable states.  So, we need to introduce a resolution timescale $\efac$ for the map to be normalizable, equivalently
 for ${\cal F}[{\cal M}]$ to have finite matrix elements in ${\cal H}$.}
  \ei

\heading{Refined criterion}

To address the first issue in two dimensions is simple\footnote{At present cannot offer an equally simple criterion
in more than two dimensions.  At present all our conclusions about quantum information for \ac{qft} in $D > 2$, are derived indirectly by considering dimensional reductions with nontrivial topology for the compact dimensions.}: We simply demand that the factorization map be a product of
operators at the separate points defining the entangling surface, rather than a superposition of products.  This is
sufficient to guarantee that $ {\cal M}\dag {\cal M}$ does not introduce long-range correlations in the image of a state in ${\cal H}$,
that were not present in the original state.

  As for the second subtlety, we resolve it in the spirit of~\cite{Ohmori:2014eia}, but with a criterion that allows for greater generality in the choice of ${\cal M}$. Like the definition of the factorization map in~\cite{Ohmori:2014eia}, we introduce a small but nonzero scale $\efac$ to characterize the time-resolution
of the factorization map.  We can at most require ${\cal F}[{\cal M}]$ to be an
 \rwa{almost}-local operator of size $\efac$, with the EOM of local operators obeying the EOM outside of a disc of size $\efac$ in the Euclidean
 geometry, centered on the boundary of subregion $A$ on the time-slice $\Sigma$.  By virtue of conformal invariance,
 if we have a factorization map with any size $\efac$, we can
 always define one of any smaller size, but we can never make
 the size precisely zero without sacrificing the finiteness of ${\cal M}$.

So our locality criterion for factorization maps is as follows:
\begin{mdframed}[%
  skipabove=.5cm,
  leftmargin=.5cm,%
  rightmargin=.5cm,%
  innertopmargin=.25cm%
  innerbottommargin=1.25cm
  ]
    In two dimensions, a(n almost-)local factorization map ${\cal M}$ of size $\efac$ is a map ${\cal M} : {\cal H}\mapsto {\cal H}\ll A
\tot {\cal H}\ll B$ such that ${\cal F}[{\cal M}]\equiv
{\cal M}\dag {\cal M}$ is a product of almost-local operators of size $\efac$ inserted at each boundary of a region $A$ of the time-slice $\Sigma \supset A$.
\end{mdframed}

\heading{Relation to other constructions of ${\cal M}$} 

It is immediately clear that either of the two definitions ${\cal M}$ used in~\cite{Ohmori:2014eia}, satisfy our almost-locality criterion.  The first construction, given
in the earlier part of~\cite{Ohmori:2014eia}, defines the factorization map ${\cal M}$ such that the almost-local operator ${\cal F}[{\cal M}]$
at each entangling point, 
  is a small hole of size $\efac$ with a local boundary condition for the quantum
  fields.  Conformally transforming to cylinder frame, the almost-local
  operator ${\cal F}[{\cal M}]$ corresponds to the state $\kket{{\cal F}}$
  on a circle of size $2\pi$, defined at radial time $\t\pr = - {\tt log}(\efac)$.
  The state at $\t\pr = - {\tt log}(\efac)$ is a boundary
  state $\kket{{\cal F}} = \kket{\tt bdy}$, satisfying the exact Cardy
  conditions in the usual sense, without needing to
  take any limit or approximation.  In this
  definition of ${\cal M}$, the first of the two in~\cite{Ohmori:2014eia},
  the boundary state $\kket{{\tt bdy}}$ is an input of the construction.  The
  boundary condition is present from the start,
without the need to take the scaling limit of traces of high powers
of the reduced density matrix $\r\ll A$.
  
In the second construction of~\cite{Ohmori:2014eia}, the authors define a factorization map ${\cal M}$ by regulating the bulk itself with a lattice regulator.  Here ${\cal M}$ is really a map between states of discrete systems
and ${\cal F}[{\cal M}]$ is realized as the almost-local operator defined by removing
lattice points in some finite region.  There is no local boundary
condition taken as an input, because there is no continuum theory at all to start with. The boundary state $\kket{{\tt bdy}}$ emerges only in the continuum limit, and only after taking high powers $\r\ll A\uu n$ of the reduced density matrix
for subregion $A$.  Remarkably, the authors manage to demonstrate
this emergence by direct calculation in examples of integrable lattice models, explicitly computing the spectrum of the Hamiltonian on the interval obtained
in the scaling limit, as the cross-channel description of the overlap
of boundary states.

Both constructions discussed in~\cite{Ohmori:2014eia}
do produce almost-local operators in the sense we mean here, in that ${\cal F}[{\cal M}]$ is almost-local of a size $\efac$ that
can be kept arbitrarily smaller than the
infrared scale,
so that ${\cal F}[{\cal M}]$ strictly commutes with degrees of freedom away from the resolved entangling region.  

Our own definition of local factorization maps is
more general, not using a lattice regulator or any UV regulator for the bulk \ac{cft} at all as does the second construction in~\cite{Ohmori:2014eia},
nor taking a boundary condition as an input as in the
first construction of that paper.

As to the construction in~\cite{Ohmori:2014eia} based
on boundary conditions, such a definition is equivalent
to a very strict condition on the quantum fields in the state $\kket{{\cal F}}$ of the radial Hilbert
space.  In this construction, the state $\kket{{\cal F}} \equiv \kket{{\tt bdy}}$
has exactly vanishing correlation between fields at any two points
on the radial slice at radial time $\t\pr = - {\tt log}(\efac)$.
This is a general property of \rwa{any}
boundary state $\kket{{\tt bdy}}$ obeying
the Cardy condition of unitarity in the cross-channel: Quantum fields at separated points have precisely vanishing correlation functions in $\kket{{\tt bdy}}$, where the expectation
value is defined as a $\lim\ll{\n\to 0}$ limit of correlation functions in the state at time $\t\pr = - {\tt log}(\efac) + \n$, that is the state
$\kket{{\tt bdy}}\ll\n \equiv \exp{- \n \cc H} \cc\kket{{\tt bdy}}$.  It is straightforward
to check this property by computing a correlation function in the dual channel, where the $\n\to 0$ limit of the expectation value is universal and dominated
by the ground state on the interval.

Our own more permissive criterion for a factorization map ${\cal M}$
allows other possibilities, in which the state $\kket{{\cal F}[{\cal M}]}$ can
have correlations on the scale of the circle itself.  It would be interesting to find a rule for determining which boundary state is recovered as the output
in the scaling limit for a given definition of ${\cal M}$.

\subsection{Renormalized tensor products $\tot$}\label{MoreSpecificDefinitionOfRenormalizedTensorProducts}

Much has been written ~\cite{Casini:2013rba} about the distinction between a precise tensor factorization of the Hilbert space
in the sense of linear algebra, versus the approximate sense in which the Hilbert space of a regulated \ac{qft} factorizes
into Hilbert spaces supported in complementary regions.  The latter notion is not always a tensor factorization in the usual
literal sense of linear algebra.  The simplest and most-analyzed example of this phenomenon~\cite{Casini:2013rba} is that of Abelian
lattice gauge theory, in which matter \ac{dof} are naturally associated to vertices of the lattice while gauge-connection \ac{dof} are naturally
associated to links.  When splitting the space into halves, the Hilbert space does not literally tensor factorize, given the usual
short-distance realization of the gauge theory.

However, as has been emphasized in ~\cite{Casini:2013rba} and later work (see for instance~\cite{Harlow:2015lma} and references therein), the nonfactorization in such cases is truly a short-distance artifact, which can be resolved
either by modifying the dynamics of the \ac{qft} at short distances, or by modifying the definition of the factorization map itself, or both.  See also~\cite{Harlow:2015lma},
in which the UV-sensitivity of the literal factorizability of the Hilbert space, is applied to the analysis of a gravity theory and used to derive striking consistency
conditions on the dynamics of Einstein-Maxwell theories \it via \em gauge-gravity duality.

Here, we give a broader notion of tensor factorization for continuum field theory, which we will denote by $\tot$, which is intended to be independent
of the UV-completion of the continuum \ac{qft}, and broad enough to encompass examples such as those considered in ~\cite{Casini:2013rba} and later work~\cite{Harlow:2015lma,1109.0036,1312.1183,1404.1391,1412.2730,1509.08478}. 

Consider a spatial slice $\Sigma$ that can be decomposed as
the (closure of the) disjoint union of $A$ and $B$, whose closures
intersect at the entangling surface ${\cal E}  = \pp A = \pp B = \bar{A} \cap \bar{B}$.  
We can denote the usual \ac{qft} Hilbert space ${\cal H}\ll\Sigma$ and the usual \ac{qft} Hilbert spaces
${\cal H}\ll{\bar{A}},{\cal H}\ll{\bar{B}}$ separately.   The usual tensor product would be defined by projection
maps from ${\cal H}\ll\Sigma$ to ${\cal H}\ll{\bar{A}},{\cal H}\ll{\bar{B}}$, but the Hilbert spaces of the QFT
on $A,B$ or their closures, are not defined without further information.  So the
notion of factorization should be extended to that of an isomorphism between ${\cal H}\ll\Sigma$ 
and ${\cal H}\ll{\bar{A}},{\cal H}\ll{\bar{B}}$ with some boundary conditions.  Such a map is
provided by the path integral with the insertion of the map-mapback operator  ${\cal F}[{\cal M}]$ as discussed in sec. \ref{LocalityCriterionPreciseDefinition}.

\section{Statuses of the converses of our results}\label{ConverseImplicationSection}

In the previous sections we have proven a series of logical implications: 
\bi
\item{$({\bf L\to TF})\uu{\ok}$:The existence of a local lattice regularization for a theory implies the existence of a tensor
factorization of the Hilbert space into Hilbert spaces supported in complementary spatial regions, modulo short distance degrees of freedom near the boundaries;}
\item{$({\bf TF\to BC})\uu{\ok}$:That the existence of such tensor factorizations automatically implies the existence of unitary, energy-preserving
boundary conditions;}
\item{$({\bf BC\to VGA})\uu{\ok}$ And, that the existence of such boundary conditions implies a vanishing gravitational anomaly.}
\ei
We have put green checkmarks on these implications to indicate that we have proven them in this paper.
One natural set of questions is, to what extent the converses of these implications are also true?  
\bi
\item{ $({\bf VGA\to BC})\uu{\blut}$
Does the vanishing of the gravitational anomaly imply the existence of unitary, energy-conserving boundary conditions?}
\item{$({\bf BC\to TF})\uu{\ok}$ Does the existence of boundary conditions, imply the existence of tensor factorizations?}
\item{$({\bf TF\to L})\uu{\ppq}$ And, does the existence of tensor factorizations imply the existence of a lattice construction?}
\ei
Let us comment briefly on the logical status of each of the three converse implications.

\heading{Does a tensor factorization always exist if a boundary condition exists?  (Yes).}

We can give an affirmative answer about the converse of one of our implications: The existence of a unitary boundary
condition does indeed always imply the existence of a tensor factorization respecting locality.  Intuitively, we can 
simply start with a state defined as an energy eigenstate at time $t < 0$, and modify the Hamiltonian for
time $t > 0$ we can impose a wall between the two separate parts of the space, regions $A$ and $B$.
This is
the correct idea but not quite rigorous because the imposition of the boundary is a singular instantaneous modification of the 
Hamiltonian, and the eigenstates of the Hamiltonian with boundary separating the two regions, may not quite overlap
with the original Hamiltonian.  As we have emphasized throughout, however, the obstructions to the construction of
tensor factorizations can be understood as infrared rather than ultraviolet obstructions, and we should
be able to get rid of the problem by resolving the instantaneous imposition of the boundary.  
We find, in Sec. \ref{ApproxFacStateConstruction}, that this can be done
simply by Euclidean time evolution for a short time $\L\uu{-1}$ in the Hilbert space of the unmodified Hamiltonian before imposing
the boundary: This defines finite transition amplitudes between the factorized Hilbert space and the original Hilbert space, defining
concretely the change of basis between the two.  
To be more specific, we considered the $t<0$-half of the anti-slit geometry in Fig. \ref{slit}, which defines a state on the full Hilbert space with a correlation that can be arbitrarily small by taking the size of the size of the anti-slit sufficiently short.
To remind ourselves, we used a conformal transformation which maps the left geometry to the right one in Fig. \ref{slit}, with the roles of time and space swapped, and simply computed the two-point function in the geometry after the transformation.
As a by-product, we derived a universal lower bound on the energetic
efficiency of disentangling the degrees of freedom supported in two complementary spatial regions of the spatial slice.
In a later subsection, we gave an example in the case of a free fermion, and then give a general construction for any
theory admitting a boundary condition.

\heading{Does a boundary condition always exist if the anomaly is nonvanishing? (Unclear but maybe.)}

Given a vanishing gravitational anomaly, can we conclude that there exist tensor factorizations respecting the locality 
of the algebra?  We have indicated this implication with a single question mark: We do not know the answer to this question, but one possible clue can be found in the nature of the Cardy--Calabrese construction~\cite{Calabrese:2004eu,Calabrese:2009ez,Calabrese:2009qy}.  In principle, a reduced density matrix is defined by its spectrum, and the information about its spectrum is encoded
in the Renyi entropies.  To the extent that the Cardy--Calabrese construction fully defines the Renyi entropies, it 
also constructs the full density matrices if all possible branched covers of the space, with branching locus at the entangling surface, are taken into account.

However, one must proceed with caution: The Cardy--Calabrese formalism gives an efficient way of \rwa{approximating} Renyi entropies at a given order, but 
it is UV-sensitive, with the cutoff entering through the ambiguity associated with the unresolved conical deficit on the singular Riemann surface, or equivalently
through the absolute normalization of the twist operator in the $n$-fold copy of the underlying \ac{cft}.  We explain this point further in Appendix \ref{sec:CardyCalabrese}.

The question of boundary conditions for non-trivial anomaly-free chiral systems has been explored recently in interesting examples~\cite{Smith:2020nuf,Smith:2020rru,Smith:2020nuf} and~\cite{Razamat:2020kyf}.

\begin{figure}
\captionsetup{singlelinecheck=off}
  \begin{tikzpicture}
    \begin{small}
      \dosserif
      \node at (0,0) {\includegraphics[width=\textwidth]{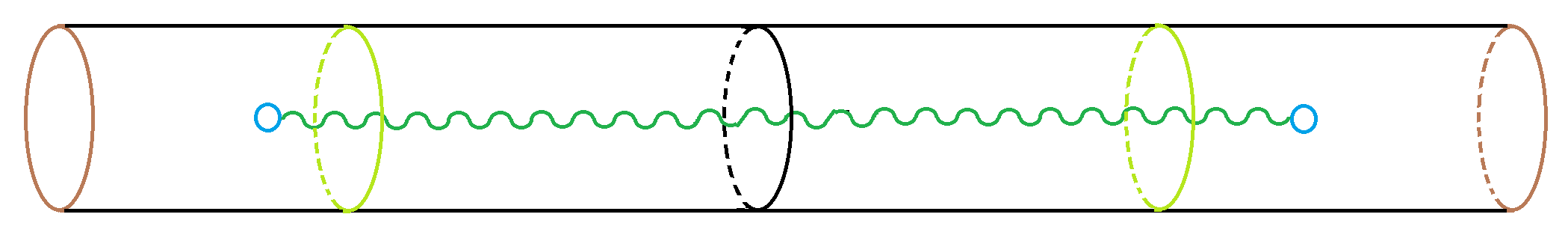}};
      \node at (0,-5) {\includegraphics[width=\textwidth]{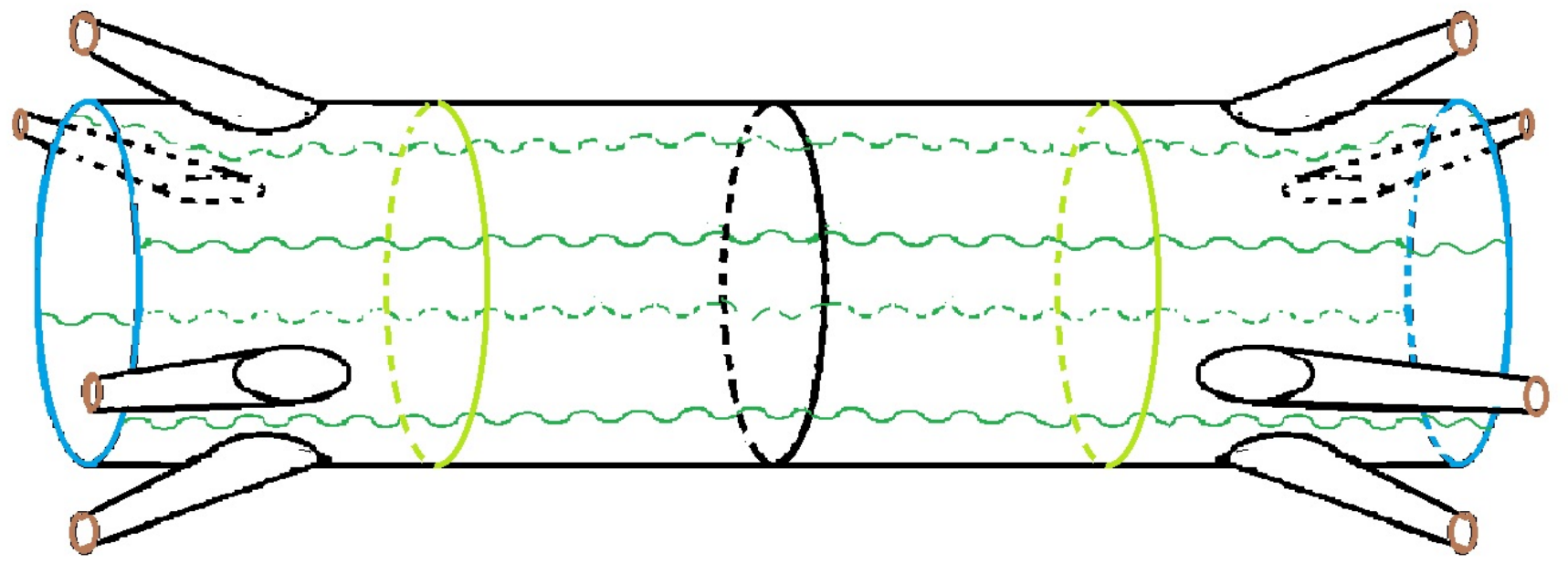}};
      \node at (0,-2) {(a)};
      \node at (0,-8) {(b)};
    \end{small}
  \end{tikzpicture}
  \caption[aaa]{
Geometries relevant for our discussion.
\begin{itemize}  
  \item[(a)] In the figure above we depict the trace ${\tt tr}(\r\uu n\ll I)$ of the n$\uth$ power of the reduced density matrix for subregion $\MacroForIOrA$, where
  the state of the full system is a thermal density matrix.  In the figure, we take $n=4$ and the subregion $\MacroForIOrA$ is an interval of finite length,
  whose boundaries are the two light blue dots.   The green squiggly line between them is a branch cut whose monodromy is a cyclic permutation
  of the four replicas of the underlying \ac{cft}.   The singularity
  of the Riemann surface in the lower figure is resolved by the finite factorization-time $\efac$, which cuts off the region containing the conical deficit and
  is represented by the finite thickness of the blue dots. 
    \item[(b)] In the lower figure we show the full Riemann surface realized as the covering space of the upper figure.   A conformal transformation
   has been performed to make the blue circle the largest scale in the geometry other than the length of the interval.
\end{itemize}   
}

  \label{fig:quotient-double}
\end{figure}

\heading{Does a lattice regulator exist if a boundary condition exists?  (Unclear but see~\cite{Wen:2013ppa,Wang:2013yta}).}

The direct implications $({\bf TF\to L})$ is a far more difficult implication than any we have considered so far.  
We wish only to observe that the general philosophy expressed in~\cite{Wen:2013ppa,Wang:2013yta} posits
that the anomaly is the only obstruction to the construction of a lattice regulator for a quantum field theory: In other words,
the program of~\cite{Wen:2013ppa,Wang:2013yta} is to establish the converse of the generalized Nielsen--Ninomiya theorem
in any dimension.  Lattice regulators for chiral systems have been constructed based on these ideas and argued to flow
to the continuum theory of interest, quite explicitly in the case of the Pythagorean fermion system~\cite{Wang:2013yta}, and under some 
plausible, well-defined assumptions in four-dimensional cases such as the $SO(10)$ chiral gauge theory or the Standard Model in~\cite{Wen:2013ppa}.

If established, this hypothesis would establish the 
equivalence of all the conditions described here, existence of lattice regulators, existence of local Hilbert spaces, existence of boundary conditions, and vanishing of the anomaly.  This is a bold and ambitious program.  It may be interesting to establish evidence in
its favor by constructing unitary boundary conditions in these cases.  This would then be equivalent to the existence of
local Hilbert spaces.  The implication $({\bf TF\to L})\uu{\ppq}$ would still be a major unbridged gap, but perhaps not
an impassable one: Tensor factorizing the space into sufficiently fine cells, under a given energy cutoff, would be tantamount
to localizing quantum information to lattice sites, with a finite number of states at each site by virtue of the cutoff.  It would then
remain to show that one could prescribe interactions between the different Hilbert spaces in such a way as to recover the original
continuum Hamiltonian in the infrared, in the limit where the cells were taken infinitely fine.

\heading{Parity}

Throughout of the main line of the logic of this section, we note briefly another
sufficient condition for the existence of boundary conditions and quantum
information.  A \ac{cft} with a parity symmetry in $D=2$
(or a \ac{qft} with a non-anomalous $\IZ\ll 2$ parity symmetry $\hat{\bf P}$ in \rwa{any} dimension) always admits an energy-preserving, unitary boundary condition, $\co(x\lrm{bdy}) = \hat{{\bf P}}
\cc \co(x\lrm{bdy}) \cc \hat{{\bf P}}\uu{-1}$ for a
point on the boundary of a spatial slice.  Then, any theory with such a symmetry automatically admits a notion of localized quantum information as well.

\section{FAQ}%
\label{sec:FAQ}

\subsection{Why doesn't the Cardy-Calabrese formalism define a tensor factorization without additional input?}%
\label{sec:CardyCalabrese}

In this paper, we have proven the nonexistence of any locally-defined notion of quantum information, for two-dimensional
theories with $C\ll L \neq C\ll R$.  

Many\footnote{Possibly all; see the discussion in sec. \ref{ConverseImplicationSection}.} \ac{cft} with $C\ll L = C\ll R$, have well-defined tensor factorizations respecting locality.  In the 
main body of the paper, we have discussed some sufficient criteria for such factorizations to exist.  When they do exist,
the traces of the n$\uth$ power of the reduced density matrix for a subregion $A$ in a given state $\kket{\Psi}$,
can often be computed most easily in the elegant formalism of~\cite{Calabrese:2004eu,Calabrese:2009ez,Calabrese:2009qy}.  In this formalism,
one promotes the \ac{cft} to an $n$-fold copy of itself, clones the state $\kket{\Psi}$ to an $n$-fold symmetric product of itself $\kket{\Psi}\uu{\otimes n}$,
and inserts twist operators permuting the copies, at the entangling surface.  Modulo singular behavior at the entangling surface,
the partition function in the $n$-fold copy with twist operators inserted,
is equivalent to the path integral of a single copy of the \ac{cft}, on an $n$-fold cover of the space, branched over the entangling surface.
It is therefore natural to try to dispense with the underlying logic of Hilbert spaces and replicas, and simply take replicated \ac{cft} with twist operator insertions,
as a \rwa{definition} of the traces $t\ll n  = {\tt tr}(\r\ll A\uu n)$ of the reduced density matrix $\r\ll A$.

While natural-seeming, this proposal is not valid: The numbers $t\ll n$ defined this way, do not have the correct properties to be traces of
any density operator $\r\ll A$.  The most obvious reason is dimensional analysis: The correlation functions $\langle \t\ll n \cc \t\ll n\dag \rangle$ of twist operators are
not dimensionless, but have an engineering dimension $({\tt length})\uu{-2\d\ll n}$ where $\d\ll n \equiv {c\over{24}}\cc (n - {1\over n})$,
while a density operator is
by definition dimensionless, having unit trace ${\tt tr}(\r\ll A) = 1$.  

\subsection{Why can't one fix up the units by arbitrarily inserting an $n$-dependent dimensional
fudge factor?}

Of course, one can try to insert an arbitrary, $n$-dependent dimensional
constant to make up the dimensions, perhaps attempting to interpret $\e\ll{\rm short}\uu{+ 2\d\ll n}$ times the correlation function as ${\tt tr}(\r\ll A\uu n)$,
where $\e\lrm{short}$ is a generic short-distance scale, such as the bulk UV distance scale $\e\ll{\textsc{uv}}$ or the factorization time-scale $\efac$.
And indeed, when a well-defined local tensor factorization exists, the \rwa{leading long-distance approximation} to
${\tt tr}(\r\ll A\uu n)$ does scale as $\e\uu{+ 2\d\ll n}\lrm{short}\cc L\uu{-2\d\ll n}$, where $L$ is the size of the subregion ~\cite{Cardy:2010zs}, where $\e\ll{\rm short}$ is the resolution scale
of the factorization map or the UV length $\L\uu{-1}\ll{\textsc{uv}}$, depending on how the factorization is defined. In order to normalize the traces of the density matrix consistently, one needs to
take $\e\lrm{short}$ finite.  For finite $\e\lrm{short}$, however, there are always subleading contributions~\cite{Cardy:2010zs}
suppressed by powers of ${\e\lrm{short}/ L}$, which one cannot discard.

\subsection{Why not just discard the subleading contributions, and use the normalized leading contributions as a definition of $\r\ll A$?}\label{WhyNotDiscardSubleadingCorrectionsFAQ}

One could try simply discarding the corrections, and interpreting the twist operator correlation functions, normalized by the UV scale,
as the traces ${\tt tr}(\r\ll A\uu n)$.  However, this can be shown easily to be inconsistent: The numbers $t\ll n \equiv (\e\lrm{short} / L)\uu{2 \d\ll n}$ simply
are not consistently interpretable as traces of powers of any fixed operator $\r\ll A$.  Any such traces must have a representation of
the form $\sum\ll k \l\ll k\uu n$, for some spectrum $\{\l\ll k\}$ that is independent of $n$.  By taking $n$ large one can see easily that 
$(\e\lrm{short} / L)\uu{2\d\ll n}$ is never of this form --
To see this, simply expand $(\e\lrm{short} / L)\uu{2\d\ll n}$ and $\sum\ll k \l\ll k\uu n$ around $n\to\infty$ and compare the two,
\begin{align}
\left(\frac{\epsilon_{\rm short}}{L}\right)^{\frac{c}{24}(n-\frac{1}{n})}
&=
\left(\frac{\epsilon_{\rm short}}{L}\right)^{\frac{c}{24}n}\times\left(1+\frac{c}{24}n\log\left(\frac{L}{\epsilon_{\rm short}}\right)+\cdots\right)
\\
\sum_{k=0}^{\dim (\mathcal{H}_A)} \lambda_{k}^n&=\lambda_0^n\times \left(1+\left(\frac{\lambda_1}{\lambda_0}\right)^n+\cdots\right)
\end{align}
which immediately tells us that the former can never reproduce the latter.
This observation was first made in~\cite{Cardy:2010zs} to derive constraints on the
subleading large-$(L/\e\lrm{short})$ corrections to the Renyi entropies, by imposing the consistency condition that a well-defined spectrum
$\{ \l\ll k \}$ of $\r\ll A$ must exist.  These corrections were reinterpreted in the large-$n$ limit in~\cite{Ohmori:2014eia}, as defining a boundary condition \it via \rm a universal recipe similar to the one in this paper.

In a \ac{cft} defined abstractly from bootstrap data, it is not automatically the case that such choices of higher-dimension operators
UV-completing the twist in a unitary way, can necessarily be made.  Indeed, in this paper we have shown that higher-energy components of the twist operator sometimes {\em{cannot}} be chosen to define UV-complete the twist in a unitary manner.  The gravitational anomaly is one obstruction to the existence of
a UV-completed factorization.  Other such obstructions may also exist, but this question is beyond the scope of this paper.

\subsection{Can the nonfactorization be cured by adding a decoupled spectator sector?}

Adding a decoupled spectator sector to the theory may cure the nonfactorization between regions, but
only at the cost of entangling the original theory
maximally with the added sector.  Such a trick was used for instance in~\cite{Iqbal:2015vka} in an attempt to define reduced density matrices for local regions in gravitationally anomalous \ac{cft}.
While this procedure is valid, it produces reduced density matrices only for the original theory and spectators combined, not for either sector separately, because the spectator sector does not remain decoupled in any finite-energy state of the factorized Hilbert space: Due to the necessity of coupling left and right-moving excitations with
short-distance interactions at the entangling surface, the result is that any state with finite eigenvalue under the modular flow, has left- and right-moving states maximally entangled as the regulator is removed.
In other words, we have simply replaced the entanglement
between regions with an entanglement between
sectors in the same region.  There is no well-defined
density matrix available for a single chirality in a single
region.

The added degrees of freedom may be useful for studying quantum information in gravitationally anomalous \ac{cft} or even for lattice regulation of such \ac{cft}, but it is essential to take into account that they are always entangled with the original degrees of freedom in any context where quantum information is localized.
They are better thought of as ``spectanglers''.

\begin{figure}
  \centering
  \begin{tikzpicture}

    \draw[-latex] (-2,2) -- (-1,2) node[above left] {time};
    \begin{scope}[thick]
    \draw[color=white, fill=black!10] (0,0) -- (2,0) -- (4,2) -- (2,2) -- cycle; 
    \draw[] (2,2) -- (0,0) -- (2,0) -- (4,2);
    \draw[dashed] (5,2) -- (1,2);
    \draw[] (-2,-2) -- (0,0) -- (2,0) -- (0,-2);
    \draw[dashed] (-3,-2) -- (1,-2);
    \draw[dashed] (0,-2) -- (-1,-3);
    \draw[dashed] (-2,-2) -- (-3,-3);
    \draw[dashed] (4,2) -- (4.5,2.5);
    \draw[dashed] (2,2) -- (2.5,2.5);

    \draw[dashed] (0,-2) -- (0,-4);
    \draw[dashed] (-2,-2) -- (-2,-4);
    \draw[] (2,0) -- (2,-2);
    \draw[] (4,2) -- (4,0);
    \draw[dashed] (0,-4) -- (0,-5);
    \draw[dashed] (-2,-4) -- (-2,-5);
    \draw[dashed] (2,-2) -- (2,-3);
    \draw[dashed] (4,0) -- (4,-1);

    \draw[dashed] (-1,0) -- (0,0);
    \draw[dashed] (2,0) -- (3,0);

    \node at (2,1) {\(A\)};
    \node at (0,-1) {\(B\)};

    \node at (0, -6.5) {(a)};
  \end{scope}

  \begin{scope}[thick, shift={(6,-.5)}]
    \draw[color=white, fill=black!10] (1,1) -- (3,1) -- (5,3) -- (3,3) -- cycle; 
    \draw[color=white, fill=black!10] (1,0) -- (1,1) -- (3,1) -- (3,-2) -- (2,-2) -- (2,0) -- cycle;

    \draw[] (3,3) -- (1,1) -- (3,1) -- (5,3);
    \draw[dashed] (6,3) -- (2,3);
    \draw[dashed] (5,3) -- (5.5,3.5);
    \draw[dashed] (3,3) -- (3.5,3.5);

    \draw[] (-2,-2) -- (0,0) -- (2,0) -- (0,-2);
    \draw[dashed] (-3,-2) -- (1,-2);
    \draw[dashed] (0,-2) -- (-1,-3);
    \draw[dashed] (-2,-2) -- (-3,-3);

    \draw[dashed] (-1,0) -- (0,0);
    \draw[dashed] (2,0) -- (3,0);
    \draw[dashed] (0,1) -- (1,1);
    \draw[dashed] (3,1) -- (4,1);

    \draw[dashed] (0,-2) -- (0,-4);
    \draw[dashed] (-2,-2) -- (-2,-4);
    \draw[] (2,0) -- (2,-2);
    \draw[] (3,1) -- (3,-2);
    \draw[] (1,1) -- (1,0);
    \draw[] (5,3) -- (5,1);

    \draw[dashed] (0,-4) -- (0,-5);
    \draw[dashed] (-2,-4) -- (-2,-5);
    \draw[dashed] (2,-2) -- (2,-3);
    \draw[dashed] (3,-2) -- (3,-3);
    \draw[dashed] (5,1) -- (5,0);

    \node at (3,2) {\(A\)};
    \node at (0,-1) {\(B\)};
    \node at (2,.5) {\(A'\)};

    \node at (0, -6) {(b)};

  \end{scope}

\end{tikzpicture}
  \caption{If the \ac{cft} is realized as edge excitations in a three-dimensional theory (a), separating the two halves of the edge introduces a new pair of half-edges (\(A'\) and \(B'\)) (b), entangled at order \(O(1)\) with the ones of interest.}
  \label{fig:new-edges}
\end{figure}
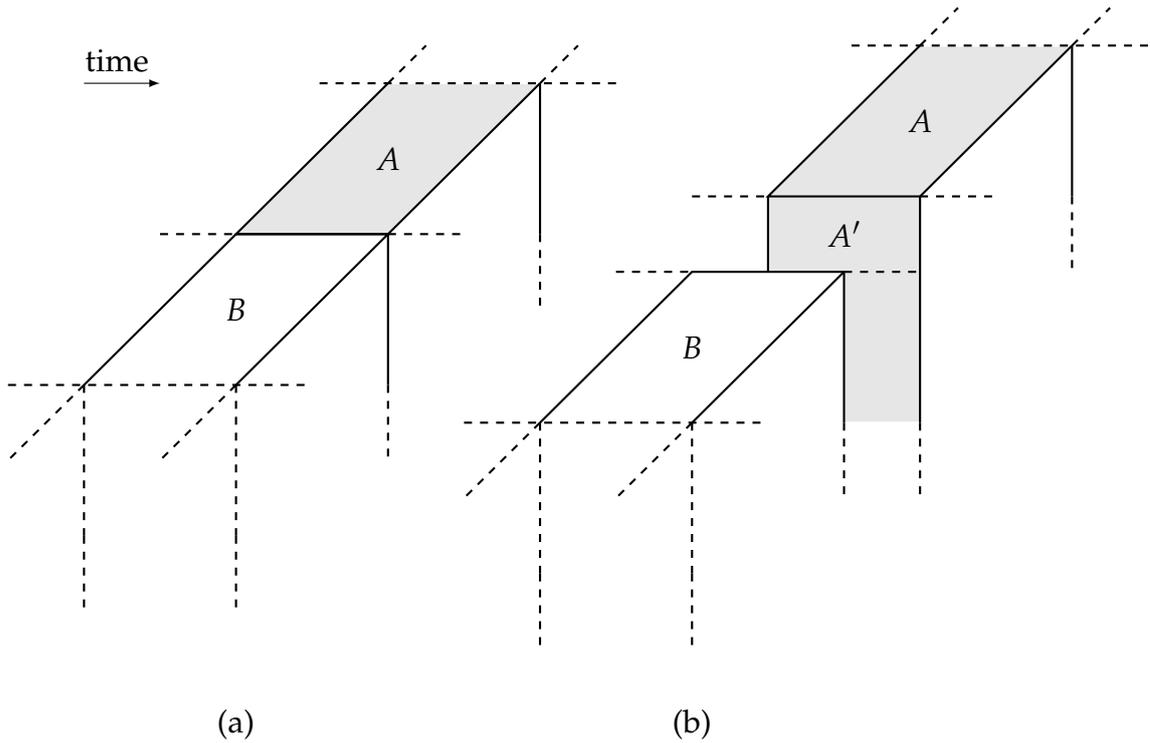

Something similar applies to the case when the anomalous \ac{cft} is realized as a set of edge excitations in a three-dimensional gapped theory (see~\cite{RevModPhys.83.1057} and references therein). 
Any attempt to physically isolate two halves of the edge of the system from one another, will simply split the bulk into two and create a new pair of half-edges supporting massless excitations in what was previously the gapped bulk, with which the previously existing massless excitations are entangled at order $O(1)$ (see Figure~\ref{fig:new-edges})

\subsection{Boundary entropy, modular flow and the Rindler Hamiltonian}

The discussion of boundary entropy illuminates one subtle point.  It is possible
to attempt to evade the no-go theorems by considering classes of regulators such
as many of those considered in~\cite{Iqbal:2015vka}, which lead to 
boundary conditions with infinite boundary entropy in the 
sense of~\cite{Affleck:1991tk}.  Some such regulators include coupling the degrees of freedom of the 2D \ac{cft} to a field theory in one or more additional noncompact direction;
another such example includes attempting to interpret the modular flow defined abstractly by Tomita--Takesaki theory~\cite{takesaki1970tomita} as a 
flow generated by a ``boost operator'' or ``Rindler Hamiltonian''~\cite{Casini:2008cr} constructed as an integral of the local Hamiltonian density.

The Tomita--Takesaki construction~\cite{takesaki1970tomita} defines a \rwa{bare} modular operator, whose finite-time modular flow is never trace-class.  In a non-anomalous theory,
the modular flow can be modified locally at distance $\e$ away from the origin and counterterms added, to obtain a renormalized modular flow which is trace-class
even in the limit $\e\to 0$.  This cannot be done in a gravitationally anomalous theory; there is simply no available local modification of the modular flow, as
exemplified by the case of purely left- or right-moving fermions or bosons in $1+1$ dimensions.  

The bare modular operator is a well-defined object in its own right, but cannot be interpreted information-theoretically absent a regularization and renormalization process
to render the modular flow a trace-class operator.  Interpreted as a density matrix, the bare modular operator of Tomita--Takesaki's construction, defines a
non-normalizable distribution which cannot be interpreted probabilistically.

The same comments apply to other constructions corresponding to boundary conditions of infinite boundary entropy, for instance coupling the two-dimensional
\ac{cft} to a noncompact three-(or higher-)dimensional \ac{qft} with massless degrees of freedom.  Such boundary conditions always have strictly infinite boundary entropy,
and do not define normalizable probability distributions.

\section{Boundary and Bulk RG Flow in Two Dimensions}
\label{BoundaryAndBulkRGFlow}

In various parts of the paper (Sections~\ref{sec:lattice-implies-boundary} and~\ref{sec:factorization-implies-boundary}) we have
appealed to the notion that a unitary but not-necessarily-conformal
bulk \ac{qft} in $D=2$ with a unitary but not-necessarily-conformal boundary condition, must flow to a \ac{cft} with conformal boundary 
condition, so
long as the bulk flows to a relativistic theory of any kind.

For a \ac{cft} in the bulk, this statement is true without
additional hypotheses: It is a standard argument
following from the Zamolodchikov $C$-theorem.
This fact about the bulk RG flow is not actually needed in our proof in section \ref{sec:quantum-info-refinement}, but we review
it because it adds valuable context to illuminate the issue of
boundary RG flow, which we do need to understand for purposes
of sections \ref{sec:lattice-implies-boundary} and \ref{sec:factorization-implies-boundary}.

\subsection{Bulk RG flow and the Zamolodchikov c-theorem}

Start by considering a non-conformal 2D theory in the bulk, away from any boundary, defined
at a scale $\L\lrm{UV}$, thought of as a cut-off quantum
field theory in the sense of the Wilsonian renormalization
group~\cite{Wilson:1971bg,Polchinski:1983gv,Wetterich:1989xg}. The cutoff theory can be in the continuum
or on the lattice, preserving the 2D relativistic symmetry or not,
with the regulator unitary or otherwise.  In our
discussion in \ref{sec:lattice-implies-boundary} we have considered a Hamiltonian lattice cutoff, which preserves unitarity exactly
and breaks relativistic symmetry at short distances.

By the logic of the renormalization group, we expect that we will
lose bulk and boundary degrees of freedom as we flow downwards in energy or equivalently view the system on longer distance scales.  Eventually
the system will reach a nontrivial fixed point,
or else lose all its degrees of freedom,
which still amounts to reaching a fixed point, the trivial/empty \ac{cft}.  

The usefulness of the Zamolodchikov $c$-theorem, and its
counterpart, the $g$-theorem, is that it makes this
intuitive picture sharp without having to follow any details of the RG flow.

\heading{Review: Proof of the Zamolodchikov $c$-theorem}

For completeness, we review the proof of the Zamolodchikov $c$-theorem below, following~\cite{polchinski1998string}.
In \(d = 2\), one can define left- and right-moving \(C \) functions that extend the conformal \(c\) coefficients to any \ac{qft}.
Following chapter 15 in~\cite{polchinski1998string} we define two functions \(C_L\) and \(C_R\) that transform along the flow according to
\begin{equation}
  \dot C_L = \dot C_R = -3/4 H
\end{equation}
where \(H\) is positive-definite for a unitary theory and vanishes if and only if the theory is conformal. Hence, \(C_L\) and \(C_R\) decrease monotonically and their difference remains constant
\begin{align}
  \dot C_L = \dot C_R & \le 0 \\
  \dot C_L - \dot C_R &= 0 .
\end{align}\label{CFlowEqs} 
This completes the proof that the left- and the right-$c$-functions are monotonic and bounded below, as well as that the difference between them are preserved under the RG flow (anomaly matching).

\heading{Unitary, relativistic \ac{qft} always flow to a \ac{cft}}

In all events, we are interested in cutoff theories in their role
as regulators for local relativistic unitary \ac{cft} in the
continuum.  It is not automatic that a cutoff \ac{qft} must
ever flow to a relativistic or unitary theory; generically they
do not.  But we are choosing our bulk cutoff theory
at $\L\ll{\textsc{uv}}$, with the couplings tuned so that the infrared limit
is relativistic and unitary.  So we can discuss the situation where
by assumption, the \ac{qft} is relativistic and unitary to
any desired approximation, by the scale $ \L\ll{ {{\rm unitary,}\atop{\rm relativistic}}}$.
\vskip.1in
At scales $\L\lesssim  \L\ll{ {{\rm unitary,}\atop{\rm relativistic}}} \muchlessthan \L\ll{\textsc{uv}}$, 
one can apply the Zamolodchikov $c$-theorem to prove
the theory must flow to a conformal fixed point.
The $C$-function $C(\L)$ is a functional on the space of
 quantum field theories and the renormalization scale $\L$,
 that extends the notion of the total central charge $C\ll L + C\ll R$ to nonconformal relativistic theories.
 $C(\L)$ decreases monotonically with the energy scale $\L$, and cannot be negative in a unitary theory.  The
$C$-theorem further proves that the bulk \ac{qft} must become
conformal when the $C$-function is stationary, $C\pr(\L) = 0$.
It follows immediately that every relativistic \ac{qft} must reach
a \ac{cft} to arbitrarily good approximation by some
scale $\L\lrm{CFT}$ with
$ \L\lrm{CFT} \muchlessthan  \L\ll{ {{\rm unitary,}\atop{\rm relativistic}}}$.

\heading{The gravitational anomaly is invariant under RG flow}

It is sometimes the case that the \ac{cft} at $\L\lrm{CFT}$
is the trivial \ac{cft}, with no degrees of freedom, $c\ll L = c\ll R = 0$.
However, we note that in the case of interest, with nonvanishing gravitational
anomaly, we can never flow to a trivial \ac{cft}; the infrared limit
will always have positive central charge.

We have referered to ``the'' $C$-functional, but
in fact the $C$-functional is defined chirally, with separate
functionals $C\ll{L,R}(\L)$ which are both
separately monotonic as a function of RG scale.  In fact the difference between the two can be shown to be \rwa{independent} of $\L$.

From this it follows that:
\bi
\item{(a) The gravitational anomaly can be defined
for any relativistic \ac{qft} as $C\ll L(\L) - C\ll R(\L)$, not just for \ac{cft};}
\item{(b) The anomaly, so defined, is independent of $\L$, even
though the two individual contributions each decrease monotonically; and}
\item{(c) A unitary relativistic \ac{qft} with nonzero gravitational
anomaly, must always flow to a nontrivial \ac{cft} with positive central
charge.}
\ei

Having reviewed these familiar facts about bulk RG flow, let us now
consider the case of boundary RG flow.

\subsection{Boundary entropy, $g$ theorem, and the hypothesis of Friedan and Konechny}\label{BoundaryRGFlowAndFKHypothesis}

Now consider the cutoff theory on the half-line $\IR\uu +$,
with a unitary but not-necessarily-conformal boundary condition.
Under some mild hypotheses, one can show that the
boundary condition must eventually flow to a conformal one,
by a sufficiently low energy scale $\L\ll{{{\rm conformal}\atop{\rm boundary }}}$.

Let us not try to track the RG flow in the energy range where
the bulk theory is still nonconformal.  In this range the boundary
dynamics are driven primarily by the RG flow of the bulk theory,
and the number of boundary degrees of freedom may increase,
rather than decrease.  We only begin following the flow
of the boundary theory at scales below $\L\lrm{CFT}$, where the
bulk \ac{qft} has flowed to a conformal point already.

At this scale $\L\lrm{CFT}$, the boundary need not be conformally
invariant in any approximation; there will in general still be nonconformal operators with coefficients whose scale may be set
by the various scales $\L\ll{\textsc{uv}}, \cc\cc  \L\ll{ {{\rm unitary,}\atop{\rm relativistic}}} , \cc\cc \L\lrm{CFT}$.
At this scale $\L\lrm{CFT}$, the boundary need not be conformally
invariant in any approximation; there will in general still be nonconformal operators with coefficients whose scale may be set
by the various scales $\L\ll{\textsc{uv}}, \cc\cc  \L\ll{ {{\rm unitary,}\atop{\rm relativistic}}} , \cc\cc \L\lrm{CFT}$.
This should intuitively still flow to a conformal boundary condition if we further go down the RG flow.

Due to the  $g$-theorem proving the monotonicity of the boundary entropy under RG flow, this intuition is automatically correct if one assumes the validity of the \emph{Friedan--Konechny hypothesis}, first discussed in~\cite{Friedan:2003yc} and its significance emphasized in~\cite{Friedan:2005dj} ); with the assumption of the
boundedness-below of the boundary entropy, it follows that every boundary theory must flow to a conformal \acl{bc}, by some sufficiently
low energy scale $\L\lrm{{{\rm conformal}\atop{\rm boundary}}}$.

\setstretch{1}
\begin{small}
  \dosserif
  \bibliography{references,references-which-cannot-be-generated-by-inspirehep}{}
  \bibliographystyle{JHEP}
\end{small}
\setstretch{1.15}

\end{document}